\documentclass[twocolumn]{aastex631}

\usepackage{amsmath, amssymb}
\usepackage{amsfonts}
\usepackage{gensymb}
\usepackage{upgreek}
\usepackage{graphics}
\usepackage{hyperref}
\usepackage{comment}

\newcommand{\swift}{{Swift}}

\begin{document}

\title{Deep Swift/UVOT Observations of GOODS-N and the Evolution of the \\ Ultraviolet Luminosity Function at $0.2<z<1.2$}

\author[0000-0002-8606-0797]{Alexander Belles}
\affiliation{Department of Astronomy \& Astrophysics, The Pennsylvania State University, University Park, PA 16802}

\author[0000-0001-6842-2371]{Caryl Gronwall}
\affiliation{Department of Astronomy \& Astrophysics, The Pennsylvania State University, University Park, PA 16802}
\affiliation{Institute for Gravitation and the Cosmos, The Pennsylvania State University, University Park, PA 16802}

\author[0000-0003-1817-3009]{Michael H. Siegel}
\affiliation{Department of Astronomy \& Astrophysics, The Pennsylvania State University, University Park, PA 16802}

\author[0000-0002-1328-0211]{Robin Ciardullo}
\affiliation{Department of Astronomy \& Astrophysics, The Pennsylvania State University, University Park, PA 16802}
\affiliation{Institute for Gravitation and the Cosmos, The Pennsylvania State University, University Park, PA 16802}

\author[0000-0002-6689-6271]{Mat J. Page}
\affiliation{University College London, Mullard Space Science Laboratory, Holmbury St. Mary, Dorking, RH5 6NT, UK}

\correspondingauthor{Alexander Belles}
\email{aub1461@psu.edu}

\begin{abstract}
    We present \swift\ Ultraviolet Optical Telescope (UVOT) observations of the deep field GOODS-N in four near-UV filters. A catalog of detected galaxies is reported, which will be used to explore galaxy evolution using ultraviolet emission. \swift/UVOT observations probe galaxies at $z \lesssim 1.5$ and combine a wide field of view with moderate spatial resolution; these data complement the wide-field observations of GALEX and the deep, high angular resolution observations by HST. Using our catalog of detected galaxies, we calculate the UV galaxy number counts as a function of apparent magnitude and compute the UV luminosity function and its evolution with redshift. From the luminosity function fits in various redshift bins, we calculate the star formation rate density as a function of redshift and find evolution consistent with past works. We explore how different assumptions such as dust attenuation corrections can dramatically change how quickly the corrected star formation rate density changes with redshift. {At these low redshifts, we find no trend between UV attenuation and redshift or absolute magnitude with significant scatter in the UV spectral slope $\beta$.} This dataset will complement the extensive observations of GOODS-N already in the literature. 
\end{abstract}


\section{Introduction} \label{sec:intro}
There have been many astrophysical breakthroughs enabled by deep field observations. The original Hubble Deep Field (HDF) found large numbers of high redshift galaxies, providing a glimpse of the early universe \citep{1996AJ....112.1335W}. Many deep field observations followed the success of HDF with larger fields of view and multiwavelength investigations. One such endeavor is CANDELS (Cosmic Assembly Near-infrared Deep Extragalactic Legacy Survey), which observed five separate fields using WFC3 and ACS on the Hubble Space Telescope (HST) \citep{2011ApJS..197...35G}. One of the CANDELS fields, GOODS-N (Great Observatories Origins Deep Survey North), covers the original HDF and has existing observations from other space based observatories \citep{2004ApJ...600L..93G, 2011A&A...533A.119E}. Here, we add deep \swift/Ultraviolet Optical Telescope (UVOT) \citep{swift_ref, UVOT_ref} observations of the GOODS-N field to the existing data in the literature. These data will facilitate a number of investigations, including the exploration of the UV galaxy number counts, the evolution of the luminosity function (LF) with redshift, and modeling of the spectral energy distributions (SEDs) of interesting sources such as Lyman Break Galaxies (LBGs). Previous work using deep UVOT imaging of the Chandra Deep Field South (CDF-S) showed the breadth of science possible by these deep near-UV observations \citep{2009ApJ...705.1462H, 2011ApJ...739...98B, 2015ApJ...808..178H}. 

Observations of galaxies using the far or near UV directly probe {recent} star formation. For galaxies at redshifts below 1.5, the rest-frame FUV can only be studied using space-based UV telescopes. HST is the highest angular resolution UV telescope and can reach depths of $m_{AB} \sim 30$ \citep{2013ApJS..209....6I} but is limited by a narrow field of view. This is problematic as galaxy surveys of the high redshift universe must be done over a large area to mitigate the effects of cosmic variance \citep{2004ApJ...600L.171S}. Wide field imaging surveys like those done by GALEX are not hindered by cosmic variance, but GALEX observations are confusion-limited due to low angular resolution \citep{2005ApJ...619L..11X}. Sitting intermediate between these extremes is \swift/UVOT which has a larger field of view ($\sim17'$) than HST and higher resolution ($\sim2.5"$) than GALEX. Additionally, UVOT provides more granular wavelength coverage in the near-UV than GALEX. The Ultraviolet Imaging Telescope (UVIT) on the AstroSat observatory occupies a similar niche, with its larger, half degree field of view, sub 2 arcsec spatial resolution, and similar aperture size (37.5 cm) compared to UVOT \citep{2012SPIE.8443E..1NK, 2017AJ....154..128T, 2023ApJS..264...40M}. Other studies have used the Optical Monitor (OM) on XMM-Newton to perform similar observations \citep{2001A&A...365L..36M, 2015A&A...574A..49A, 2021MNRAS.506..473P}. The UV emission of galaxies at these redshifts remain valuable as they can be paired with observations across the electromagnetic spectrum to understand the total energy budget of galaxies. 

While ultraviolet data of galaxies in this redshift range are scarce, there are extensive datasets available at longer wavelengths. The original Great Observatories Origins Deep Survey combined IR HST imaging, ground-based optical data, Spitzer mid-IR photometry and deep Chandra X-ray observations \citep{2003AJ....126..539A, 2004ApJ...600L..93G}. The CANDELS survey provides UV through IR data over 5 fields of which {deep multiwavelength HST observations exist for $\sim125$ square arcminutes of the two GOODS fields} \citep{2011ApJS..197...35G, 2011ApJS..197...36K}. Later HST surveys, {such as HDUV \citep{2018ApJS..237...12O} and UVCANDELS \citep{2024RNAAS...8...26W}, provided additional deeper UV imaging of the GOODS fields, adding to the wealth of data available for these fields}. Grism and photometric data also exist from the 3D-HST survey \citep{2014ApJS..214...24S, 2016ApJS..225...27M}. 

These multiwavelength data are essential to pair with UV observations. As dust preferentially affects UV radiation, multiwavelength data can be used to model both the stellar populations and the dust properties of the galaxy \citep[for details on modeling the spectral energy distribution of galaxies, see][]{2011Ap&SS.331....1W, 2013ARA&A..51..393C}. Additionally, more data across the electromagnetic spectrum improves the precision of photometric redshift estimates, which are necessary to calculate luminosity from observed fluxes \citep{2019NatAs...3..212S}. 

Results from GALEX and HST are essential to our understanding of the UV evolution of galaxies. Observations using GALEX provided a wide-field sample of the UV emission of galaxies at $z\sim1$, which previously could only be observed with HST. These observations enabled measurements of the galaxy population over a large area as well as over a large fraction of cosmic time. \citet{2005ApJ...619L..11X} studied the FUV and NUV galaxy number counts down to 24.5 AB mag, finding results consistent with an evolving LF. \citet{2005ApJ...619L..15W} determined the LF in the local universe ($z<0.1$), while \citet{2005ApJ...619L..43A} and \citet{2005ApJ...619L..47S} studied the evolution of the LF and star formation rate density out to $z\sim1.2$.

Data from HST have been used to study the faintest galaxies. \citet{2010ApJ...725L.150O} traced the evolution of the UV LF in the GOODS-S field at $0.75<z<2.5$ using HST and found that at $z\sim1$, the faint end slope of the luminosity function was steeper than what was found in the local universe. Recent work by \citet{2023arXiv231115664S} used data from the UVCANDELS project to calculate the evolution of the UV LF in a larger area than past works. Additionally, Astrosat UVIT has produced deep field imaging in the UV to study stellar populations over this redshift range \citep{2023arXiv231001903B, 2023ApJ...946...90M, 2023ApJS..264...40M}.

Beyond LF science, these observations allow detailed study of the UV spectral energy distributions of high redshift galaxies, which are highly sensitive to the effects of dust. \citet{2011ApJ...726L...7O} found that high redshift LBGs are consistent with the IRX-$\beta$ relationship originally derived in \citet{1999ApJ...521...64M}. More work has continued in studying the evolution in the UV spectral slope of galaxies, which is indicative of evolution in the stellar populations as well as dust attenuation properties \citep{2014ApJ...793L...5K}. \citet{2018ApJ...853...56R} probed fainter galaxies at $z\sim2$, finding that their IRX-$\beta$ relation is consistent with a Small Magellanic Cloud-like extinction law \citep{2003ApJ...594..279G}. The finer wavelength resolution provided by \swift/UVOT will be used to understand dust properties at $z\sim1$.

Previous UVOT work using observations of the Chandra Deep Field South (CDF-S) showed its ability to complement the multiwavelength and other UV data in the literature. \citet{2009ApJ...705.1462H} showed that the UV number counts were consistent with an evolving LF. \citet{2015ApJ...808..178H} later quantified the evolution in the Schechter function parameters with redshift \citep{1976ApJ...203..297S}. These previous works suffered from incompleteness at the faint end of the luminosity function, meaning the faint end slope $\alpha$ was unconstrained \citep{2015ApJ...808..178H}. Recent works by \citet{2021MNRAS.506..473P} and \citet{2022MNRAS.511.4882S} using OM, on which UVOT was designed, studied the UV LF in the ROSAT 13h deep field and the CDF-S respectively, with more recent work done on the COSMOS field \citep{2022arXiv221200215S}. \citet{2022MNRAS.511.4882S} found characteristic magnitudes that were about half a magnitude less luminous than those found by \citet{2015ApJ...808..178H} for the same field and were able to constrain the faint end slope of the Schechter function. {In this work, we will construct a catalog of UVOT detected sources in GOODS-N that will be used in conjunction with data from the literature to further study the redshift evolution of star-forming galaxies. With the additional near-UV coverage provided by UVOT, we will explore correcting the effects of dust attenuation.}

Currently, observations by JWST are pushing the boundaries of our knowledge by finding star-forming galaxies at $z\sim10$ \citep{2023NatAs.tmp...66C, 2023NatAs.tmp...67R}. For galaxies at this extreme redshift, JWST probes the rest-frame UV and optical emission. Observations of the rest-frame infrared, which can be used to constrain the effects of dust, are not possible for these objects. JWST is also finding massive red galaxies at $z\sim8$, raising questions about our cosmological models and theories of galaxy formation \citet{2023Natur.616..266L}. As we reach to study the first generations of galaxies, it is also imperative to continue to expand our knowledge at lower redshifts where data are more plentiful. These lower redshift results will serve as a important comparison when studying the very early universe.

The paper is laid out as follows: we present our UVOT data in \S\ref{sec:data}, provide our results on the UV number counts and UV luminosity function in Section \ref{sec:results}. Finally, we discuss our results in Section \ref{sec:disc} and present conclusions in Section \ref{sec:conclusion}. Throughout, the AB magnitude system is used \citep{1983ApJ...266..713O} along with a flat $\Lambda$CDM cosmology with $h=0.7$, $\Omega_m=0.3$ and $\Omega_\Lambda=0.7$.

\section{Data}
\label{sec:data}

\swift\ observations of GOODS-N ($\alpha_{J2000}$, $\delta_{J2000}$ = 189.20\degree, 62.2314\degree) occurred daily throughout 2021. Data were taken using 4 near-UV filters (UVW2, UVM2, UVW1, and $u$), as well as contemporaneous X-ray observations taken with XRT as part of a Swift Cycle 17 Guest Investigator program (Prop. No.: 1720039). In this Section, we summarize the observations and discuss the data processing procedure.  

\subsection{UVOT}

\begin{figure}
    \centering
    \includegraphics[width=0.45\textwidth]{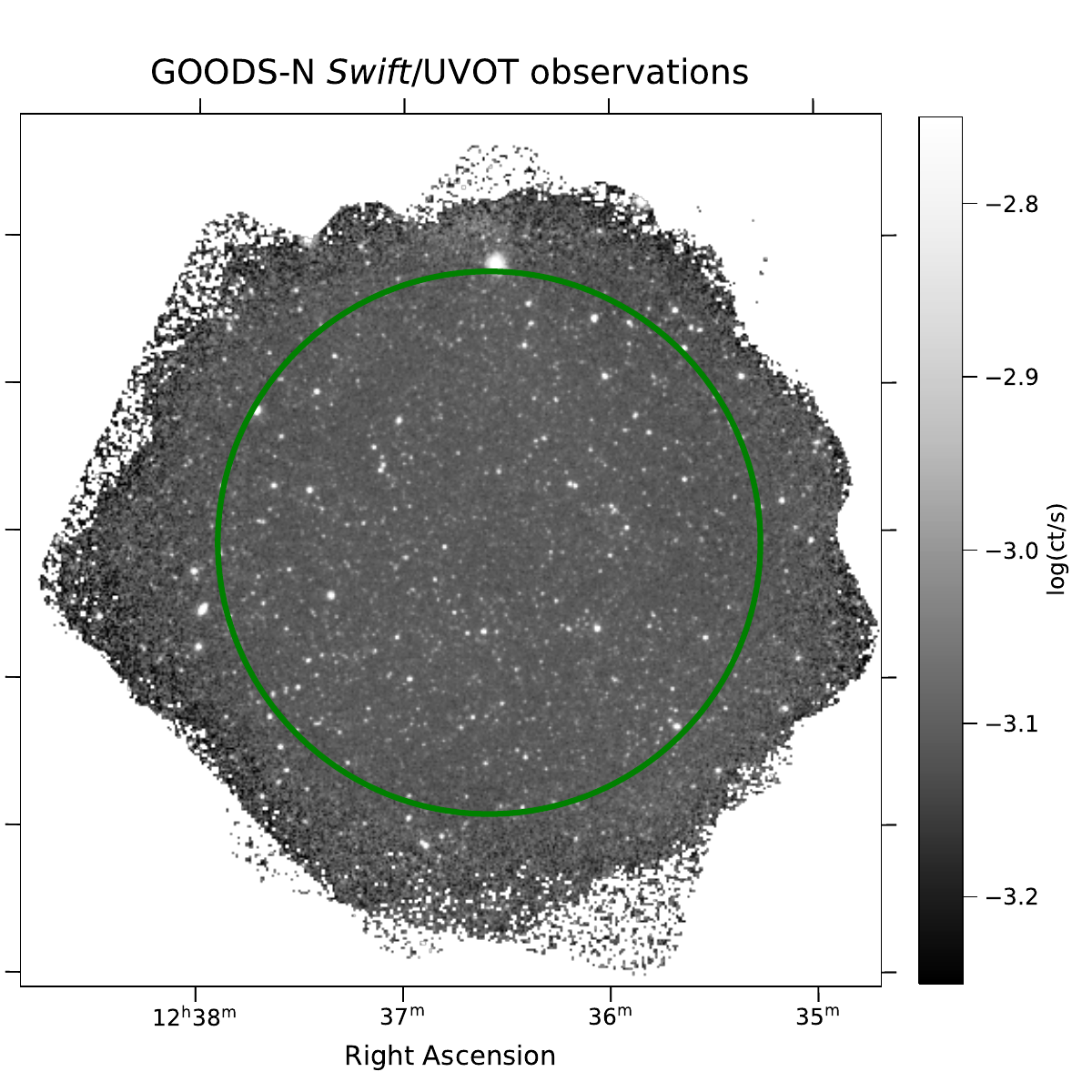}
    \caption{GOODS-N as observed using the \swift/UVOT UVM2 filter. The UVM2 image is shown due to its low background compared to the other filters. The green circle marks the region where we detected the sources presented here. Sources outside this region were excluded due to the lower effective exposure time at the edge of the mosaicked field.}
    \label{fig:m2_footprint}
\end{figure}

As the longest continuous observation by \swift\ is limited to 1.8 ks, many observations of GOODS-N were required to achieve the desired depth. Observations were taken daily from April 1st to December 24th, 2021 in one of the four near UV filters. In total, the field was observed for a total of 80, 78, 80, and 73 ks in the UVW2, UVM2, UVW1, and $u$ filters respectively. {The central wavelengths for these filters are 1967, 2260, 2629, and 3469 \AA, and the quoted FWHM of the PSF are 2.92'', 2.45'', 2.37'', and 2.37''.} The exposures for each filter were combined into a single mosaic. UVOT has a field of view of 17 arcminutes and pointing accuracy on the order of 1 arcminute. Combined with roll angle differences, the actual total observed field is slightly larger than expected from the 17 arcminute FOV.

Data were processed in two ways. For each method, we first correct for bad pixels and large scale sensitivity variations across the detector \citep{2014AJ....148..131S}. For our first processing method, we sum the counts images and the exposure maps separately using the Heasoft provided tool \texttt{uvotimsum}. We then create a total count rate image by dividing the summed counts image by the total exposure map. Due to a variable background between exposures, this method of combining the data led to a non-uniform background in the final count rate image. This was particularly an issue in the $u$ band images, where the background count rate varied by up to a factor of ten between observations, likely due to scattered light from the Earth limb.

To mitigate this, we use the \texttt{astropy} \citep{astropy:2013, astropy:2018, astropy:2022} coordinated package \texttt{reproject} \citep{reproject} to average the count rate images of individual exposures with background matching. The background matching is done by calculating the measured count rate differences of overlapping regions and adding a constant offset to each image in order to minimize these differences. While the average correction is constrained to be zero, this procedure may add a small offset to our final mosaic. However, this procedure leads to a more uniform background in the final count rate image. 

We calculate photometry using both stacking methods. We verify that the standard UVOT point source photometry tool \texttt{uvotsource} gives comparable fluxes for both stacking methods. However, we find a more uniform background when combining the $u$ band images using count rate averaging, with no negative impact on the other filters. Therefore, the mean averaged count rate image is used as our mosaicked image for all filters. In Figure~\ref{fig:m2_footprint}, we show the UVM2 mosaic image and highlight the outer regions where data were excluded due to lower effective exposure times.

After the final science images were created, sources were identified and photometry measured with Source Extractor \citep{1996A&AS..117..393B}. First, we create a metaimage that is used for detecting sources and determining the aperture shape and size. Our metaimage is the {sum of the count rate images for the UVW2, UVM2, and UVW1 filters. Compared to using a single band, this ensures that we detect sources even if they are not seen in each filter.} The $u$ band data are excluded from the metaimage due to the overall higher background level even after our averaging procedure. To best measure the brightest sources as well as identify the faintest objects, we employ a two step approach for Source Extractor \citep[see][]{2013ApJS..206...10G, 2013ApJS..207...24G}. For the first step (``cold" mode), we identify the brightest sources without aggressively deblending, using similar configuration settings as those described in \citet{2009ApJ...705.1462H} for CDF-S data. The second mode (``hot" mode) is configured to detect the faintest possible sources. Running SExtractor in cold mode would not split a clumpy galaxy into multiple sources while hot mode would. After running SExtractor in both modes, we combine the sources into one catalog.  

All sources detected using the cold mode settings are included in the catalog. For the sources detected in the hot mode, we exclude all hot mode identified sources {that fall within the Kron aperture} of a cold source. {This prevents double counting sources and also prevents bright sources from needlessly being split into multiple sources.} 


We use the \texttt{MAG\_AUTO} values from SExtractor, which is based on \citet{1980ApJS...43..305K}. \texttt{MAG\_AUTO} uses a flexible elliptical aperture based on the observed light distribution for each detected source. The zero-points from \citet{2010MNRAS.406.1687B} were used, along with the most recent time dependent sensitivity loss correction.\footnote{\url{https://heasarc.gsfc.nasa.gov/docs/heasarc/caldb/swift/docs/uvot/uvotcaldb\_throughput\_06.pdf}} As the zero-points are defined for a 5 arcsecond aperture, we calculate the radius of the circular aperture equal in area to the Kron aperture and use the curve of growth for the different UVOT bands to correct the fluxes \citep[see ][]{2019MNRAS.486..743D}.  

We test that the photometry using SExtractor is accurate by confirming that that \texttt{uvotsource} photometry using a 5 arcsecond aperture agrees with the SExtractor values. We note that the distribution of the difference between the aperture and SExtractor magnitudes is centered around zero with a long tail towards the SExtractor magnitudes being brighter by up to 0.5 mag in some cases. At the same time, the errors on the SExtractor \texttt{MAG\_AUTO} values were systematically larger. 

As UVOT is a photon counting device, coincidence loss occurs when two photons arrive within the 11 ms frametime. However, analyses of the coincidence loss as a function of count rate in \citet{2008MNRAS.383..627P} and \citet{2010MNRAS.406.1687B} shows that for the low count rates seen here, the error introduced due to coincidence loss is negligible so we do not apply any coincidence loss correction. Other works such as \citet{2009ApJ...705.1462H} and \citet{2011ApJ...739...98B} also found the coincidence loss correction to be on the order of 1\%.  

\subsection{XRT}

While UVOT drives our main science, XRT observations of the field occurred simultaneously and can be used to identify X-ray sources such as active galactic nuclei (AGN). However, as this field has also been observed by the Chandra X-ray Observatory for 2 Ms \citep{2016ApJS..224...15X}, deeper X-ray observations exist in the literature. After cross-matching our catalog with the catalog from \citet{2014ApJS..215...27Y}, we separate out X-ray sources and stellar contaminants.  

\subsection{Literature Photometry}

With the UVOT photometry presented here, we cross match our detected galaxies with catalogs from previous works. We use the catalog from \citet{2014ApJS..215...27Y} to cross match with our UVOT detected sources to get photometric redshifts, spectroscopic redshifts when available, object classification (X-ray source or star), as well as photometry in the optical through infrared.

Other catalogs providing photometry and redshift estimates of objects in this field exist. For example, \citet{2019ApJS..243...22B} provides an updated catalog of H-band detected sources. Near-IR data can also be found in \citet{2019ApJ...871..233H}. These other sources of photometry only provide partial coverage of the UVOT observed field of view. Since \citet{2014ApJS..215...27Y} covers the entire area imaged by UVOT, it allows the use of self consistent redshift estimates. 

To minimize possible source confusion, we set a 1.5 arcsecond tolerance when matching UVOT detected sources to the \citet{2014ApJS..215...27Y} catalog. After matching, 56 objects are classified as X-ray sources. \citet{2014ApJS..215...27Y} state that the vast majority of X-ray sources are AGN so they are excluded here as they are assumed to not be typical star-forming galaxies. 

For our final catalog, roughly 60\% of the objects have spectroscopic redshifts. Of those with only photometric redshifts, greater than 90\% of those sources have high quality photometric redshift estimates \citep[$Q_z<1$,][]{2008ApJ...686.1503B}. {\citet{2014ApJS..215...27Y} quote a normalized median absolute deviation of 0.026 for non X-ray sources at $z<1$. They also claim a 3.9\% outlier percentage, where an outlier is defined as $\lvert z\rvert/(1+z_{spec})<0.15$. Following the advice of \citet{2014ApJS..215...27Y}, we make a photometric redshift cut by requiring $Q_z<1$. }

Table~\ref{tab:detect} contains the UVOT catalog as described. In total, 1171 sources were identified with matching counterparts within 1.5 arcseconds of source in the \citet{2014ApJS..215...27Y} catalog. {After removing poor photometric redshift estimates and stellar/X-ray sources, we are left with 1011 sources.} 

\startlongtable
\begin{deluxetable*}{cccccccccc}
\tablecaption{Catalog of UVOT detected sources in GOODS-N \label{tab:detect}}
\tablewidth{\textwidth}
\tablecolumns{6}
\tablehead{\colhead{$\alpha$ (deg)} & \colhead{$\delta$ (deg)} & \colhead{UVW2 mag} & \colhead{UVW2 err} &
\colhead{UVM2 mag} & \colhead{UVM2 err} & \colhead{UVW1 mag} & \colhead{UVW1 err} & \colhead{u mag} & \colhead{u err}}
\startdata
189.1813886 & 62.3951872 & 23.54 & 0.24 & 24.11 & 0.48 & 23.77 & 0.50 & 23.93 & 0.88 \\
189.1994436 & 62.3946238 & 23.37 & 0.24 & 23.37 & 0.29 & 22.73 & 0.23 & 22.18 & 0.21 \\
189.1669418 & 62.3924013 & 24.25 & 0.37 & 23.26 & 0.18 & 22.95 & 0.19 & 22.76 & 0.24 \\
189.228319 & 62.3895898 & 22.90 & 0.21 & 22.94 & 0.26 & 22.63 & 0.28 & 22.98 & 0.59 \\
189.1163935 & 62.3884952 & 23.78 & 0.39 & 23.24 & 0.28 & 22.82 & 0.27 & 22.60 & 0.34 \\
189.1711512 & 62.3884962 & 23.60 & 0.30 & 23.62 & 0.36 & 23.37 & 0.41 & 22.91 & 0.41 \\
189.1524967 & 62.3848727 & 23.04 & 0.22 & 22.76 & 0.20 & 22.77 & 0.29 & 22.67 & 0.40 \\
189.0225393 & 62.3845391 & 23.55 & 0.24 & 23.65 & 0.31 & 23.25 & 0.31 & 22.61 & 0.26 \\
\enddata
\tablecomments{Table 1 is published in its entirety in the machine-readable format. A portion is shown here for guidance regarding its form and content.}
\end{deluxetable*}


\section{Results}
\label{sec:results}

\subsection{Bias}\label{subsec:bias}

Before we can analyze our data, we have to understand its limitations and biases. First, our survey is flux-limited, meaning the data suffer from Malmquist bias, where brighter objects are detected out to farther distances. This means that intrinsically luminous sources can appear to be more numerous as they can be detected in a larger volume than dimmer sources. To account for this bias, we model our completeness as a function of apparent magnitude, i.e. determining the fraction of sources at a given apparent magnitude that can be detected. We calculate this by injecting artificial sources of known flux into the final mosaicked and detection images, running our detection algorithm, and recording if the artificial source was detected. This technique is common throughout the literature \citep[for example,][]{1995ApJ...449L.105S, 2009ApJ...705.1462H, 2021MNRAS.506..473P, 2022MNRAS.511.4882S}. 

{We assume that these galaxies are point sources, described using a Gaussian profile with a FWHM equal to the PSF for the given UVOT filter. \citet{2021MNRAS.506..473P} found that XMM-Newton OM detected galaxies at $z>0.6$ were point sources. As UVOT is based on and strongly influenced by OM, we assume that galaxies at the same redshift and magnitude are point-like. Thus we need to confirm that the lowest redshift galaxies at $z\sim0.2$ are point-like. To do so, we use the Kron radius, which is used to determine the flux of the galaxies, as a measure of their spatial extent. We find that, within the redshift bins $0.2<z<0.4$, $0.4<z<0.6$, $0.6<z<0.8$, and $0.8<z<1.2$, the median Kron aperture radius is relatively constant at around 5.5 arcseconds. We see a trend where the aperture size increases for fainter objects. We attribute this to the source count rate approaching the sky background level. This same increasing of aperture size for low count rates happens for the injected artificial sources. As a result, the signal to noise drops for the faintest objects.} 

{To calculate the completeness as a function of apparent magnitude, we estimate the fraction of recovered sources at discrete magnitude points. As the completeness drops rapidly at a certain level, our intervals are not evenly spaced, with smaller steps in regions where the completeness changes rapidly. At each discrete apparent magnitude point, we generate a single artificial source with a count rate corresponding to that apparent magnitude. To mimic the noise characteristics of our images, we apply Poisson noise based on the number of detected counts. The artificial source is added to both the detection image and the science image. When adding the artificial source to our metaimage, we scale its count rate by a factor of three. As the metaimage is the sum of the count rate images for the UVW2, UVM2, and UVW1 filters, this is roughly equivalent to assuming a constant spectrum in $f_\nu$. After this source is added to the detection image, we run our detection pipeline and record whether the artificial source is detected at its known position. This is repeated 500 times for each discrete apparent magnitude and then for each of the 4 UVOT filters.}

{For each UVOT filter, we} detect $\sim98\%$ of sources at the bright end and our completeness levels off at $2-4\%$ at the faint end. This implies both our confusion bias and spurious sources percentage is on the order of a few percent. This low level of confusion bias is a clear advantage over surveys by GALEX which can have source confusion at the $\sim20\%$ level \citep{2009ApJ...697.1410L}.

{Another possible source of error is the inclusion of interlopers such as stars or spurious sources. The purity of our sample is based on the purity of the \citet{2014ApJS..215...27Y} catalog as we require each UVOT detected source to have also been detected in \citet{2014ApJS..215...27Y}. We use their source classification to remove stars and X-ray objects from our sample. They find that for sources with spectroscopy, only 0.4\% of galaxies were misclassified as stars and $\sim10\%$ of stars were misclassified as galaxies. However, these misclassifications were primarily bright sources or blended with a nearby object.}

\begin{figure}
    \centering
    \includegraphics[width=0.45\textwidth]{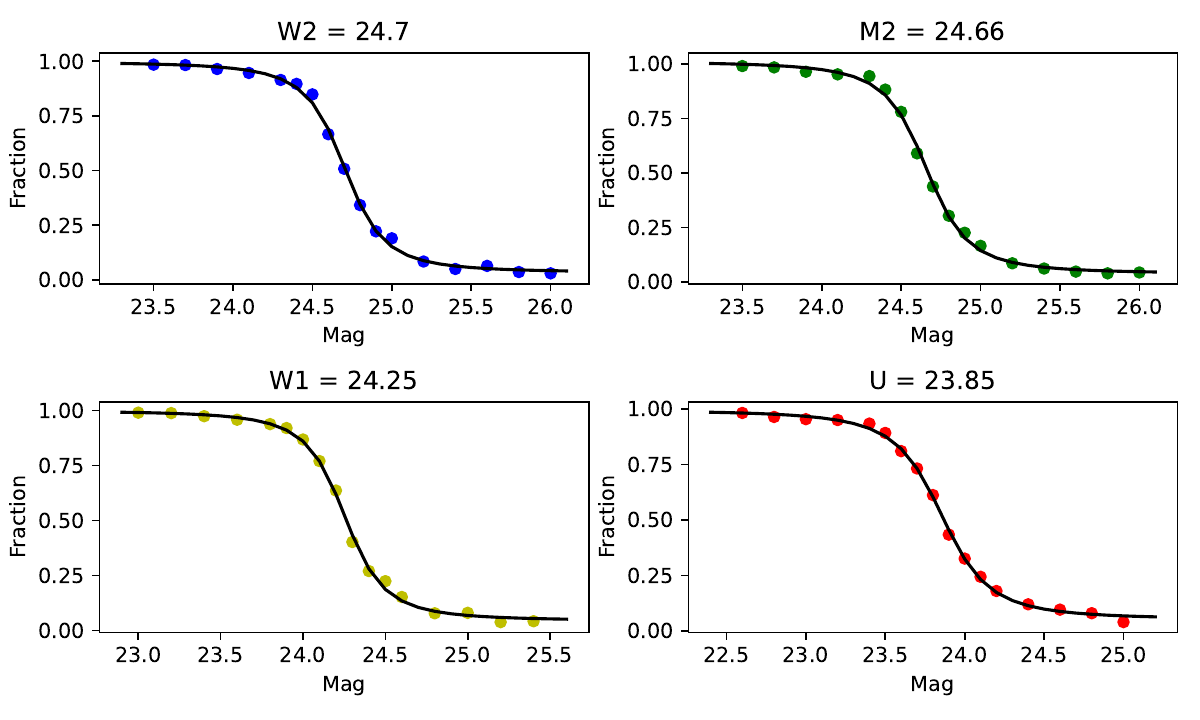}
    \caption{Completeness curves for the four UVOT bands. Detectability is a complex function of exposure time, filter transmission curve, {surface brightness,} and detection algorithm. We artificially inject point sources of {known flux} into our images and determine if they are recovered by SExtractor and at what magnitude. This is repeated to build up statistics on the recovery fraction. The completeness curves are show as well as the magnitude at which our observations are 50\% complete. }
    \label{fig:completeness}
\end{figure}

{Using the discrete measurements of the completeness fraction as a function of apparent magnitude}, we fit the data with the function defined in \citet{1995AJ....109.1044F}, 

\begin{equation}
    C(f) = \frac{1}{2} \frac{1+\xi \log(f/f_{50})}{\sqrt{1+\xi \log(f/f_{50})^2}},
\end{equation}
\noindent where $\xi$ is a constant and $f_{50}$ is the flux level at 50\% completeness. Converting $f_{50}$ to magnitudes, we find that our 50\% completeness levels for the UVW2, UVM2, UVW1, and $u$ band are 24.7, 24.66, 24.25, and 23.85 mag. The results of our completeness simulations can be seen in Figure~\ref{fig:completeness} and the Fleming curve provides an excellent fit. 

{We perform another test to determine to what extent SExtractor detected sources are impacted by source size. We repeat our detection process where our injected sources have a FWHM equal to twice the PSF. This was only done for the UVM2 filter. We find that the 50\% completeness level drops by approximately 0.8 mag, due to the lower surface brightness. We also see that the median Kron aperture size increases. The median aperture size of the point sources was 4.5 arcseconds compared to 7.5 for the extended sources added in this test. This shows that the SExtractor dynamic aperture sizing can detect extended sources by using a larger aperture albeit to a lesser depth. As the average size of the sources detected in our images is comparable to the Gaussian with FHWM equal to the PSF, we conclude that, by and large, the detected sources are compact and not extended. Nonetheless, in the lowest redshift bin ($0.2<z<0.4$) it is possible that we are overestimating the completeness at the faintest magnitudes.}

Another source of bias was described in \citet{1913MNRAS..73..359E}. This so-called Eddington bias is when measurement errors scatter sources into brighter bins. This can affect the luminosity function at the bright end, where the shape of the luminosity function is changing rapidly and only depends on a handful of objects. Here, we only discuss the observed luminosity function without an Eddington bias correction.

The last source of potential error is cosmic variance. Cosmic variance is the inherent statistical uncertainty when observing only one small part of the universe. Due to random statistical fluctuations, observations of one area of sky may be significantly different than another. As a result, the true error on a measurement is larger than the Poisson error. \citet{2004ApJ...600L.171S} quantified the effects of cosmic variance in the GOODS fields. Additionally, an online calculator is available from \citet{2008ApJ...676..767T} to estimate the cosmic variance for given survey parameters. Similar to \citet{2015ApJ...808..178H}, we find that the contribution of cosmic variance is $\sim2\times$ the Poisson error. As cosmic variance is a systematic effect in density, we do not incorporate its contribution into our quoted error bars.  

\subsection{Number Counts}

We first look at the number counts of galaxies in our four near-UV bands. We note that there is a known overdensity at $z\sim0.5$ and $z\sim0.9$ in the HDF-N, which is encompassed by the field observed here \citep{2000ApJ...538...29C}. 

\begin{figure}
    \centering
    \includegraphics[width=0.48\textwidth]{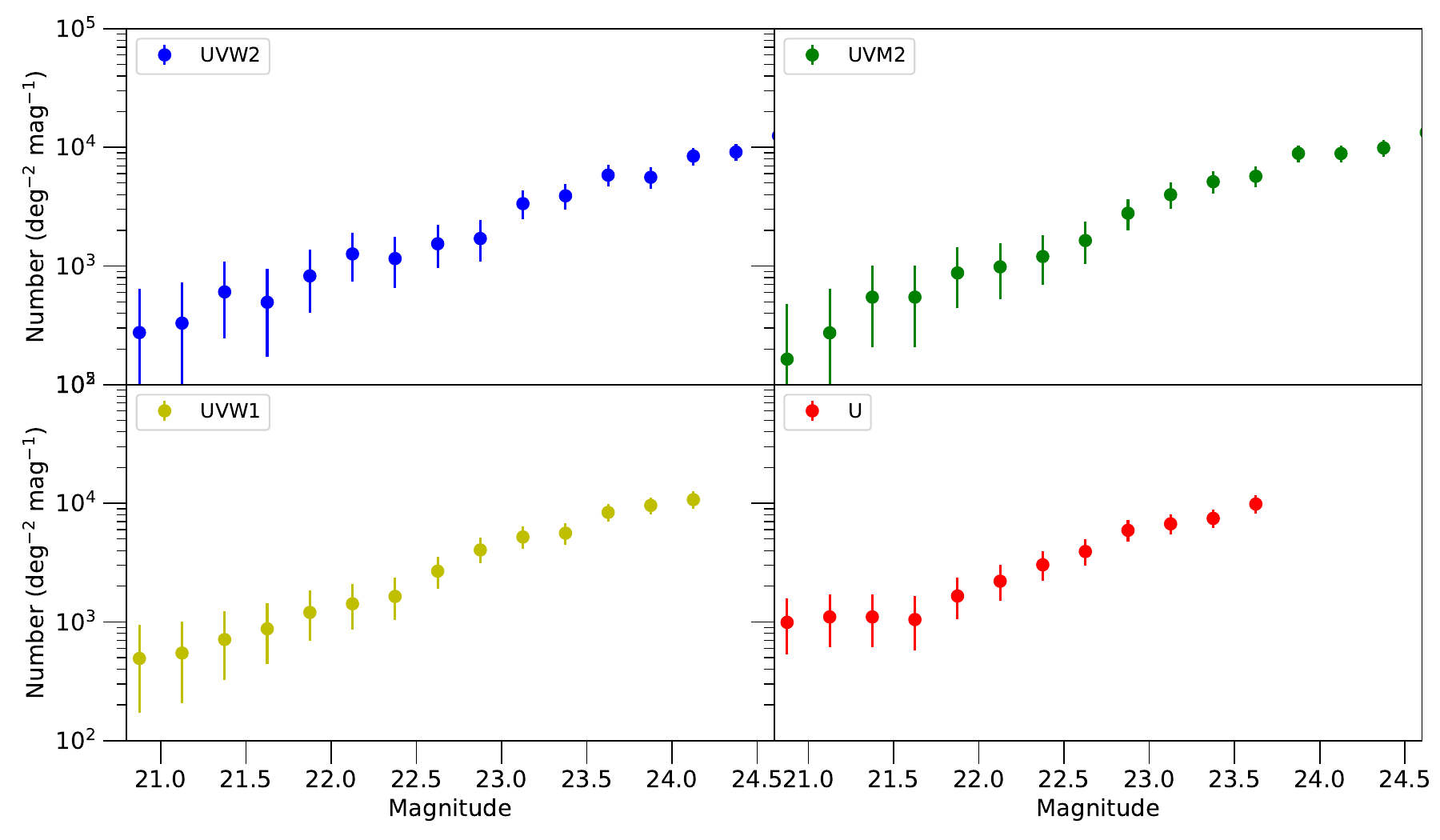}
    \caption{Galaxy number counts as a function of apparent magnitude using \swift/UVOT {down to the 50\% completeness limit}. We bin our sample in steps of 0.25 mag and calculate the number of galaxies detected per square degree per magnitude and corrected for completeness. The error bars are the Poisson noise associated with each bin \citep{1986ApJ...303..336G}. We do not increase the error bars to compensate for the effects of cosmic variance.}
    \label{fig:uvot_number_count}
\end{figure}

We bin our sample of detected galaxies into intervals of width 0.25 mag. For each bin, the number count is the total number of galaxies at that apparent magnitude divided by the survey area and magnitude bin width. This is finally normalized based on the completion fraction for each bin. The error on the number density of sources is the Poisson error as determined using \citet{1986ApJ...303..336G}. Our number counts for the four UVOT filters are shown in Figure~\ref{fig:uvot_number_count} {down to the 50\% completeness limit.}. The tabulated values are listed in Table \ref{tab:nc}. 

\startlongtable
\begin{deluxetable*}{cccccc}
\tablecaption{Number count data for each of the 4 UVOT filters\label{tab:nc}}
\tablewidth{\textwidth}
\tablecolumns{7}
\tabletypesize{\scriptsize}
\tablehead{\colhead{Bin center} & \colhead{Number count}  & \colhead{Lower limit} & \colhead{Upper limit} &\colhead{$N_{raw}$} & \colhead{Completeness} \\ \colhead{(AB Mag)} & \colhead{(deg$^{-2}$~mag$^{-1}$ )} & & & & }
\startdata
20.875 & 274.978 & 156.495 & 460.570 & 5 & 0.984 \\
21.125 & 329.973 & 199.372 & 526.659 & 6 & 0.984 \\
21.375 & 604.951 & 425.603 & 847.532 & 11 & 0.984 \\
21.625 & 494.960 & 333.342 & 720.666 & 9 & 0.984 \\
21.875 & 824.933 & 614.554 & 1097.374 & 15 & 0.984 \\
22.125 & 1264.897 & 1003.265 & 1587.241 & 23 & 0.984 \\
22.375 & 1154.906 & 905.100 & 1465.688 & 21 & 0.984 \\
22.625 & 1539.874 & 1250.785 & 1889.143 & 28 & 0.984 \\
22.875 & 1704.861 & 1400.484 & 2069.161 & 31 & 0.984 \\
23.125 & 3354.726 & 2926.500 & 3841.465 & 61 & 0.984 \\
23.375 & 3904.681 & 3442.488 & 4425.130 & 71 & 0.984 \\
23.625 & 5836.939 & 5270.995 & 6460.621 & 106 & 0.983 \\
23.875 & 5600.578 & 5041.556 & 6218.403 & 100 & 0.966 \\
24.125 & 8444.788 & 7749.151 & 9200.241 & 147 & 0.942 \\
24.375 & 9134.446 & 8394.447 & 9936.976 & 152 & 0.900 \\
24.625 & 12524.756 & 11485.959 & 13653.515 & 145 & 0.627 \\
24.875 & 23192.393 & 20964.528 & 25645.327 & 108 & 0.252 \\
\hline
20.875 & 163.987 & 75.015 & 322.992 & 3 & 0.990 \\
21.125 & 273.311 & 155.546 & 457.779 & 5 & 0.990 \\
21.375 & 546.622 & 376.944 & 779.544 & 10 & 0.990 \\
21.625 & 546.622 & 376.944 & 779.544 & 10 & 0.990 \\
21.875 & 874.595 & 658.466 & 1152.190 & 16 & 0.990 \\
22.125 & 983.920 & 754.384 & 1274.535 & 18 & 0.990 \\
22.375 & 1202.569 & 948.331 & 1517.277 & 22 & 0.990 \\
22.625 & 1639.866 & 1342.313 & 1997.060 & 30 & 0.990 \\
22.875 & 2787.772 & 2398.822 & 3235.207 & 51 & 0.990 \\
23.125 & 3990.341 & 3524.491 & 4514.050 & 73 & 0.990 \\
23.375 & 5138.247 & 4609.321 & 5724.653 & 94 & 0.990 \\
23.625 & 5706.485 & 5147.914 & 6322.614 & 104 & 0.986 \\
23.875 & 8902.615 & 8197.414 & 9666.027 & 159 & 0.966 \\
24.125 & 8877.004 & 8167.118 & 9646.071 & 156 & 0.951 \\
24.375 & 9888.530 & 9117.237 & 10722.472 & 164 & 0.897 \\
24.625 & 13332.825 & 12191.100 & 14576.786 & 136 & 0.552 \\
24.875 & 15650.535 & 13797.998 & 17736.570 & 71 & 0.246 \\
\hline
20.875 & 491.960 & 331.322 & 716.299 & 9 & 0.990 \\
21.125 & 546.622 & 376.944 & 779.544 & 10 & 0.990 \\
21.375 & 710.609 & 516.314 & 967.105 & 13 & 0.990 \\
21.625 & 874.595 & 658.466 & 1152.190 & 16 & 0.990 \\
21.875 & 1202.569 & 948.331 & 1517.277 & 22 & 0.990 \\
22.125 & 1421.217 & 1144.474 & 1757.969 & 26 & 0.990 \\
22.375 & 1639.866 & 1342.313 & 1997.060 & 30 & 0.990 \\
22.625 & 2678.448 & 2297.257 & 3118.201 & 49 & 0.990 \\
22.875 & 4045.003 & 3575.957 & 4571.886 & 74 & 0.990 \\
23.125 & 5199.474 & 4667.059 & 5789.427 & 95 & 0.989 \\
23.375 & 5601.510 & 5045.161 & 6216.078 & 101 & 0.976 \\
23.625 & 8382.110 & 7693.967 & 9129.216 & 148 & 0.956 \\
23.875 & 9599.736 & 8850.968 & 10409.322 & 164 & 0.924 \\
24.125 & 10727.597 & 9840.900 & 11690.809 & 146 & 0.737 \\
24.375 & 23753.705 & 21696.864 & 25996.884 & 133 & 0.303 \\
\hline 
20.875 & 991.935 & 760.530 & 1284.918 & 18 & 0.982 \\
21.125 & 1102.150 & 857.976 & 1407.572 & 20 & 0.982 \\
21.375 & 1102.150 & 857.976 & 1407.572 & 20 & 0.982 \\
21.625 & 1047.043 & 809.168 & 1346.324 & 19 & 0.982 \\
21.875 & 1653.226 & 1353.248 & 2013.330 & 30 & 0.982 \\
22.125 & 2204.301 & 1857.381 & 2610.677 & 40 & 0.982 \\
22.375 & 3030.913 & 2623.600 & 3497.046 & 55 & 0.982 \\
22.625 & 3921.619 & 3457.421 & 4444.326 & 71 & 0.980 \\
22.875 & 5917.351 & 5340.894 & 6552.911 & 105 & 0.960 \\
23.125 & 6711.128 & 6094.288 & 7387.459 & 118 & 0.951 \\
23.375 & 7458.237 & 6802.518 & 8174.316 & 129 & 0.936 \\
23.625 & 9857.867 & 9037.434 & 10749.609 & 144 & 0.791 \\
23.875 & 10630.856 & 9536.526 & 11844.109 & 94 & 0.479 \\
\hline
\enddata 
\tablecomments{Data are shown in order for UVW2, UVM2, UVW1, and $u$. The 1 $\sigma$ upper and lower limits are determined using \citet{1986ApJ...303..336G} and we do not correct for cosmic variance.}
\end{deluxetable*}

We compare our number counts to other studies in the literature. We focus on comparing the UVM2 number counts found here for GOODS-N to the UVM2 number counts in CDF-S from \citet{2009ApJ...705.1462H}, the GALEX NUV all-sky number counts from \citet{2005ApJ...619L..11X}, and the HST FUV number counts from \citet{2006AJ....132..853T}. Our focus is on the UVM2 results as the filter does not suffer from a red leak like the UVW2 and UVW1 filters \citep{2010ApJ...721.1608B, 2014AJ....148..131S} and is most similar to GALEX NUV, minimizing the effects of any assumptions made when applying a color correction to compare the two. 

{In order to accurately compare the observed UVM2 number counts to that of other filters, specifically the HST FUV number counts, we must determine the color difference between pairs of filters. We assume that a power law is a reasonable approximation for the spectral energy distribution (SED) in the UV. For each source in our sample, we calculate the slope $\beta$ that best fits $F(\lambda) \propto \lambda^\beta$ using the UVOT photometry. This is the slope of the observed UV spectrum and not the true rest-frame UV slope. We look at the distribution of the power law slopes and find the median is $-0.754$ and a mean of $-0.473$, implying a long tail towards larger values of $\beta$. We use the median $\beta=-0.754$ and calculate the expected UVM2-NUV color for this representative source.}

{We find that for $\beta=-0.754$, the UVM2-NUV color is $-0.018$.} This slight color difference is smaller than the {Galactic} extinction correction. Assuming a reddening of $E(B-V)\sim0.01$ \citep{1998ApJ...500..525S, 2011ApJ...737..103S}, a \citet{1999PASP..111...63F} extinction law, $R_V=3.1$, and a flat input spectrum, the correction for the UVM2 filter is 0.09 mag. Changing the flat input spectrum assumption to a spectrum with UV slope $\beta\sim-0.75$ does not meaningfully change the {MW extinction} correction. 

To compare with the \citet{2006AJ....132..853T} FUV number counts, the color correction will be non-negligible and require an assumption of the input spectrum. {We assume a power law spectrum of $\beta=-0.754$ and find the difference between the HST F150LP and UVOT UVM2 filter to be -0.274. This correction is applied to the number counts from \citet{2006AJ....132..853T}.}

\begin{figure}
    \centering
    \includegraphics[width=0.5\textwidth]{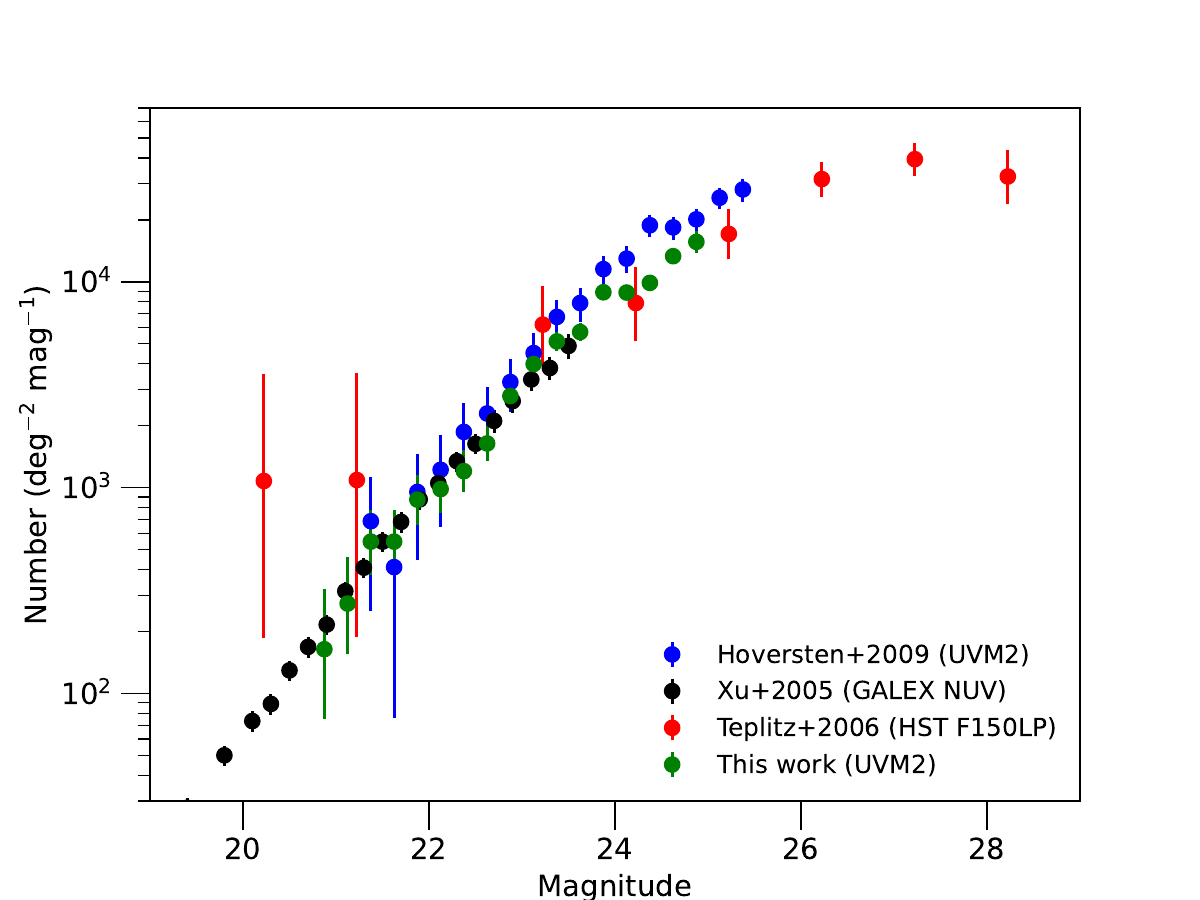}
    \caption{The UVM2 number counts (green) compared to GALEX NUV counts from \citet{2005ApJ...619L..11X} (black), UVM2 number counts of CDF-S from \citet{2009ApJ...705.1462H} (blue), and HST FUV number counts from \citet{2006AJ....132..853T} (red). We see good agreement and differences can be attributed to survey area and exposure times. The corrections for {MW extinction} are minimal ($\sim0.1$ mag) as is the color difference between UVM2 and GALEX NUV. The correction to compare UVM2 and the HST values depends on the assumed spectrum of the observed source. {A correction of $-0.274$ mag, based on an power law spectrum with slope $-0.754$, is applied to the HST values.} } 
    \label{fig:num_count_comp}
\end{figure}

Figure~\ref{fig:num_count_comp} shows the UV number counts here compared to prior works and there is good agreement at the $\sim10\%$ level. Our values here have larger errors than the \citet{2005ApJ...619L..11X} values due to the different survey sizes \citep[266 arcminutes here versus many square degrees in][]{2005ApJ...619L..11X} but our measurements go a magnitude deeper. The CDF-S values from \citet{2009ApJ...705.1462H} go slightly deeper than the work presented here due to greater exposure time (80 ks here vs $\sim$130 ks).

While the more astrophysically interesting quantity is the luminosity function, the number counts can be used to understand the galaxy population when other photometry is not available or the redshifts of objects are unknown. The number counts can be used to test evolution in the galaxy population as was done in \citet{2009ApJ...705.1462H}. As the redshifts for these objects have {already been measured or estimated in \citet{2014ApJS..215...27Y}}, we now turn our attention to the UV luminosity function.

\subsection{UV Luminosity Function}

The luminosity function (LF) is the density of galaxies per luminosity interval as a function of luminosity. The UV luminosity function can be used to trace the evolution of the star formation rate density (SFRD) as the UV luminosity is a direct probe of recent star formation \citep{2014ARA&A..52..415M}. However, the SFRD as determined by UV luminosity is subject to high uncertainty due to the scaling relation from luminosity to star formation rate and the correction for dust attenuation. Often the IR is used in tandem, as composite SFR indicators are more precise and do not suffer from dust attenuation \citep{2012ARA&A..50..531K}.  

Here, we provide our measurements of the UV LF in the redshift range $0.2<z<1.2$ to build on previous work done by GALEX \citep{2005ApJ...619L..43A} and more recent works by \swift/UVOT \citep{2015ApJ...808..178H}, XMM-Newton OM \citep{2021MNRAS.506..473P, 2022MNRAS.511.4882S, 2022arXiv221200215S}, Astrosat UVIT \citep{2023arXiv231001903B}, and HST \citep{2023arXiv231115664S}.

First, we {convert the observed UVM2 apparent magnitude of a source to an absolute magnitude using the measured redshift and the estimated K-correction from} the code \texttt{kcorrect} \citep[version 5.0.1b]{2007AJ....133..734B}. The K-correction is determined for each galaxy using the observed SED. {For each observed object, we utilize the UVOT photometry; the \citet{2014ApJS..215...27Y} photometry calculated using ground based UBVRI, z', and HK imaging from \citet{2004AJ....127..180C}, JH band imaging from \citet{2010ApJS..186...94K}, Ks band imaging from \citet{2010ApJS..187..251W}, and Spitzer IRAC data from \citet{2013ApJ...769...80A}; and the redshift value from \citet{2014ApJS..215...27Y}, of which over half are spectroscopic.}

{The absolute magnitude and redshift values are the data used to calculated the UV LF.} We fit the UV LF using two methods, the traditional $V_\text{max}$ method and a more flexible Maximum Likelihood Estimation (MLE) approach. We will discuss how the LF is calculated for each method, their strengths and weaknesses, and the results from both approaches.  

We utilize the \citet{1976ApJ...203..297S} functional form when fitting the LF:
\begin{align}
\begin{split}
    \phi(M) dM = 0.4 \log(10) \phi^* \left( 10^{0.4(M^* - M)} \right)^{\alpha+1} \\
    \times \exp(-10^{0.4 (M^* - M)}) dM, 
\end{split}
\end{align}
where $\phi(M)$ is the number density of galaxies of magnitude $M$, $\phi^*$ is the normalization, $M^*$ is the characteristic magnitude at the exponential cutoff, and $\alpha$ is the faint-end slope. 

\begin{figure}
    \centering
    \includegraphics[width=0.45\textwidth]{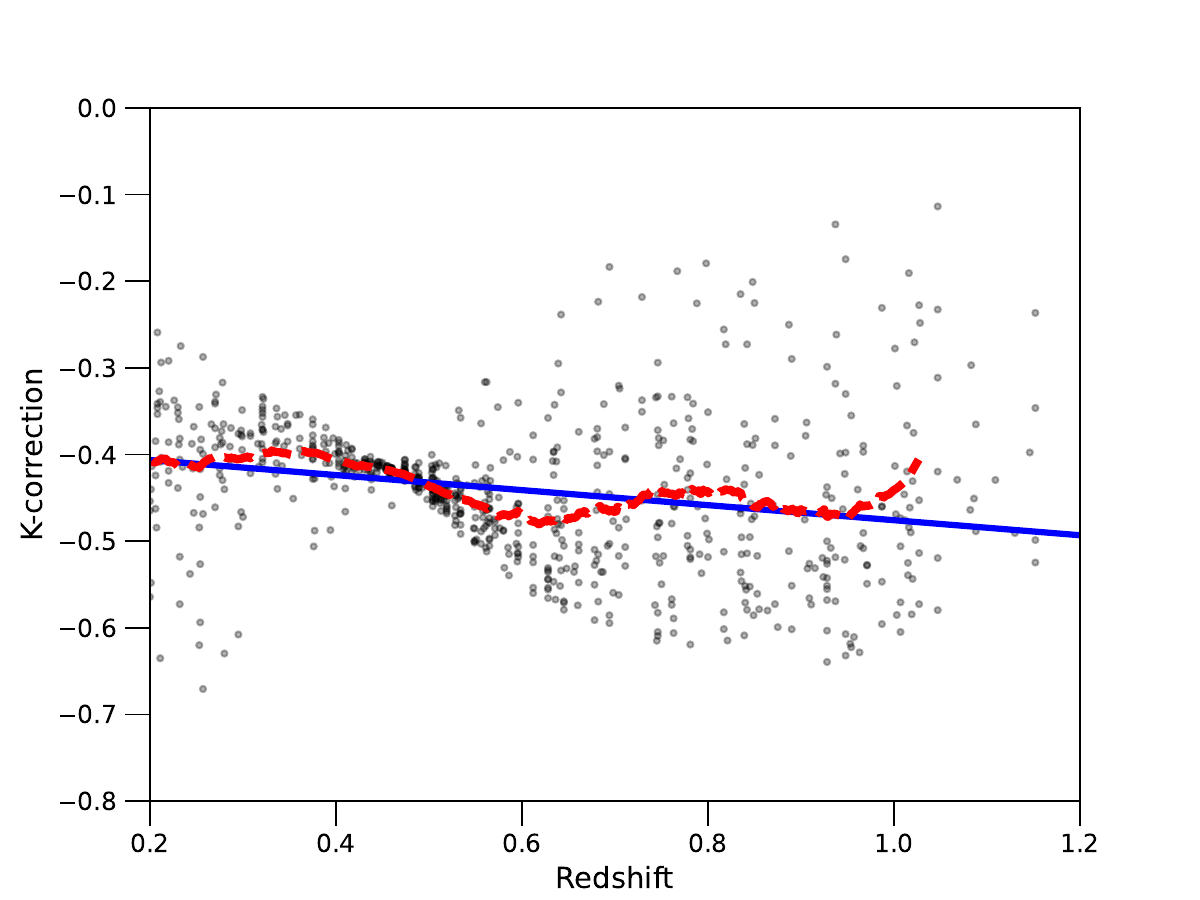}
    \caption{{The K-correction as a function of redshift for galaxies above the 50\% completeness threshold. We see that the K-correction shows a slight redshift dependence with scatter. We fit a linear function (blue line), which is a good fit to the running average (red dashed line). This regression is used in the calculation of the UV LF.}}
    \label{fig:kcorr_z}
\end{figure}

To minimize uncertainty due to incompleteness, only galaxies above the 50\% completeness limit are used when fitting the UV LF. {Figure~\ref{fig:kcorr_z} shows the calculated K-correction as a function of redshift for galaxies above the 50\% completeness limit. We see that the K-correction is only a weak function with redshift but with some scatter. In the construction of our UV LF, we assume a weak linear function with redshift to describe the K-correction. }

After our completeness cut, we calculate the LF using 795 objects. 

\subsubsection{$V_\text{max}$ Method}

Our first approach at fitting the UV LF is using the $V_\text{max}$ method introduced in \citet{1968ApJ...151..393S}. As this is a flux-limited survey, our completeness affects our ability to detect faint sources at high redshift. To account for this, the $V_\text{max}$ method relies on calculating the maximum volume in which a source could be observed multiplied by the completeness for that source magnitude. That is:

\begin{align}
    \phi(M) dM = \sum_i \frac{1}{C_i V_{max, i}} 
\end{align}
where $C$ is the completeness fraction as calculated from the \citet{1995AJ....109.1044F} function, and $V_\text{max}$ is the maximum volume in which the object could be detected by our survey. The LF is calculated with binned data to get discrete estimates of $\phi(M)$ by summing over the $i$ in the magnitude bin. As the FUV absolute magnitudes are calculated using the observed UVM2 apparent magnitude, we utilize the UVM2 completeness curve. We also perform bootstrap resampling, similar to \citet{2015ApJ...808..178H}, to calculate errors on the discrete LF points. These discrete points are then fit with a Schechter function using a Levenberg–Marquardt least squares algorithm. 

\subsubsection{MLE}

One downside of the $V_\text{max}$ method is the use of binned data. By changing the binning, the best fit parameters can vary. A MLE approach is more general and takes full advantage of the unbinned data and can be modified to incorporate photometric uncertainties \citep{2021MNRAS.506..473P}. Here, we use the likelihood function outlined in \citet{2013ApJ...769...83C} but note that other likelihood functions are used in the literature \citep{2021MNRAS.506..473P}. One advantage of the likelihood function from \citet{2013ApJ...769...83C} is the ability to jointly fit all three Schechter parameters compared to other approaches where the normalization is determined by the number of observed sources. 

We briefly outline the likelihood function used here. For an observed luminosity function $\phi(L) \Delta L$, we can say the expected number of galaxies in the bin $L$ to $L + \Delta L$ is $\lambda = \phi(L) \Delta L \Delta V$, where $\Delta V$ is the volume observed. As galaxies are being counted, the probability that you observe $n$ galaxies is given by the Poisson distribution,

\begin{align*}
    p(n) = \frac{\lambda_i^n e^{-\lambda_i}}{n!}.
\end{align*}
Extended to the whole sample, the probability of seeing $n_i$ galaxies in the bin $L_i$ to $L_i + \Delta L$ is 

\begin{align*}
    P = \prod_i p(n_i) = \prod_i \frac{\lambda_i^n \exp(-\lambda_i) }{n!}.
\end{align*}
In the limit where $\Delta V$ and $\Delta L$ go to $dL$ and $dV$, we can say that in each bin, there is either 0 or 1 galaxy.  

\begin{align*}
    P &= \prod_\text{bins with 0} \exp(-\lambda_i) \prod_\text{bins with 1} \lambda_i \exp(-\lambda_i) \\
    &= \prod_\text{all bins} \exp(-\lambda_i) \prod_{i=1}^N \lambda_i \\
    &= \exp(-\sum_{\text{all bins}} \phi(L_i) dL dV) \prod_i^N \phi(L_i) dL dV,
\end{align*}
{where $N$ is the total number of sources.}

The sum in the above equation can be re-written as an integral. When taking the log of the likelihood {and dropping constant terms}, we are left with

\begin{align}
    \ln P = - \int_{z_1}^{z_2} \int_{L_{min}(z)}^{\infty} \phi(L) dL \frac{dV}{dz} dz + \sum_i^N \ln(\phi(L_i)),   
\end{align}
{where $L_\text{min}(z)$ is the minimum luminosity that our sample is sensitive to. It is the luminosity corresponding to the 50\% completeness limit at a given redshift, using the K-correction as a function of redshift described above.}

\begin{figure*}
    \centering
    \includegraphics[width=0.85\textwidth]{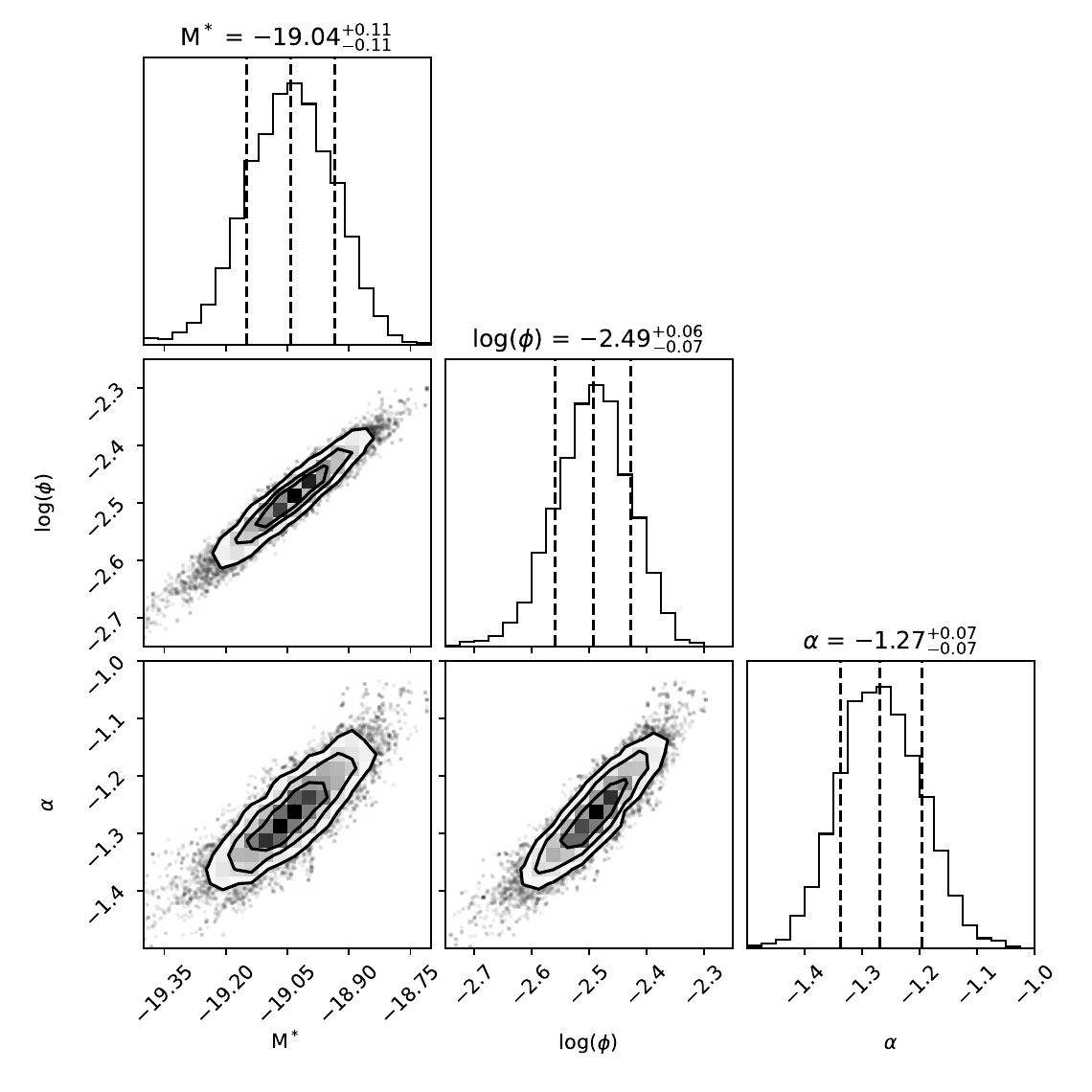}
    \caption{The diagnostic corner plot shows the Schechter function parameter estimates for the entire sample of galaxies across the whole redshift range. This diagnostic plot shows the covariances between the different parameters. However, the one dimensional posteriors are well fit with a Gaussian.}
    \label{fig:corner}
\end{figure*}

We use the above likelihood function to estimate parameters that best fit the observed luminosity function. In order to understand covariances in the data and get error estimates, we perform Markov Chain Monte Carlo (MCMC) sampling using \texttt{emcee} \citep{2013PASP..125..306F}. We first determine the parameters that minimize $-\ln P$ using standard minimization techniques. This initial estimate is used as the starting point for the walkers. By starting the walkers in the region of highest likelihood, we reduce the burn-in time. The posterior distributions we get from the MCMC sampling give us median values for the parameters as well as errors on the estimated value. An example of our 2D posteriors when determining the UV LF for the entire sample are shown in Figure~\ref{fig:corner}. 

To get information regarding evolution with redshift, we subdivide our sample into four redshift bins and calculate the LF in each. The redshift bins are $0.2<z<0.4$, $0.4<z<0.6$, $0.6<z<0.8$, and $0.8<z<1.2$. Our best fit Schechter parameters and errors are given in Table~\ref{tab:params}. 

For the full sample, both the MLE and $V_\text{max}$ results are given. In each redshift bin, we provide the MLE values. Previous UV LF work using UVOT \citep{2015ApJ...808..178H} was unable to constrain $\alpha$ in any redshift bin. Here, we are able to estimate $\alpha$ in the two lowest redshift bins and give the results with $\alpha$ as a free parameter. We also report the MLE values when we fix $\alpha$ using the values determined in \citet{2005ApJ...619L..43A}. We note that other results from this redshift range \citep[e.g.][]{2010ApJ...725L.150O, 2020MNRAS.494.1894M, 2021MNRAS.506..473P, 2022arXiv221200215S} find $\alpha$ estimates with a large dispersion, making it difficult to determine any evolution in $\alpha$ over this redshift range. Recent work by \citet{2023arXiv231001903B} finds evolution in $\alpha$ using UVIT. 

\begin{figure*}
    \centering
    \includegraphics[width=0.9\textwidth]{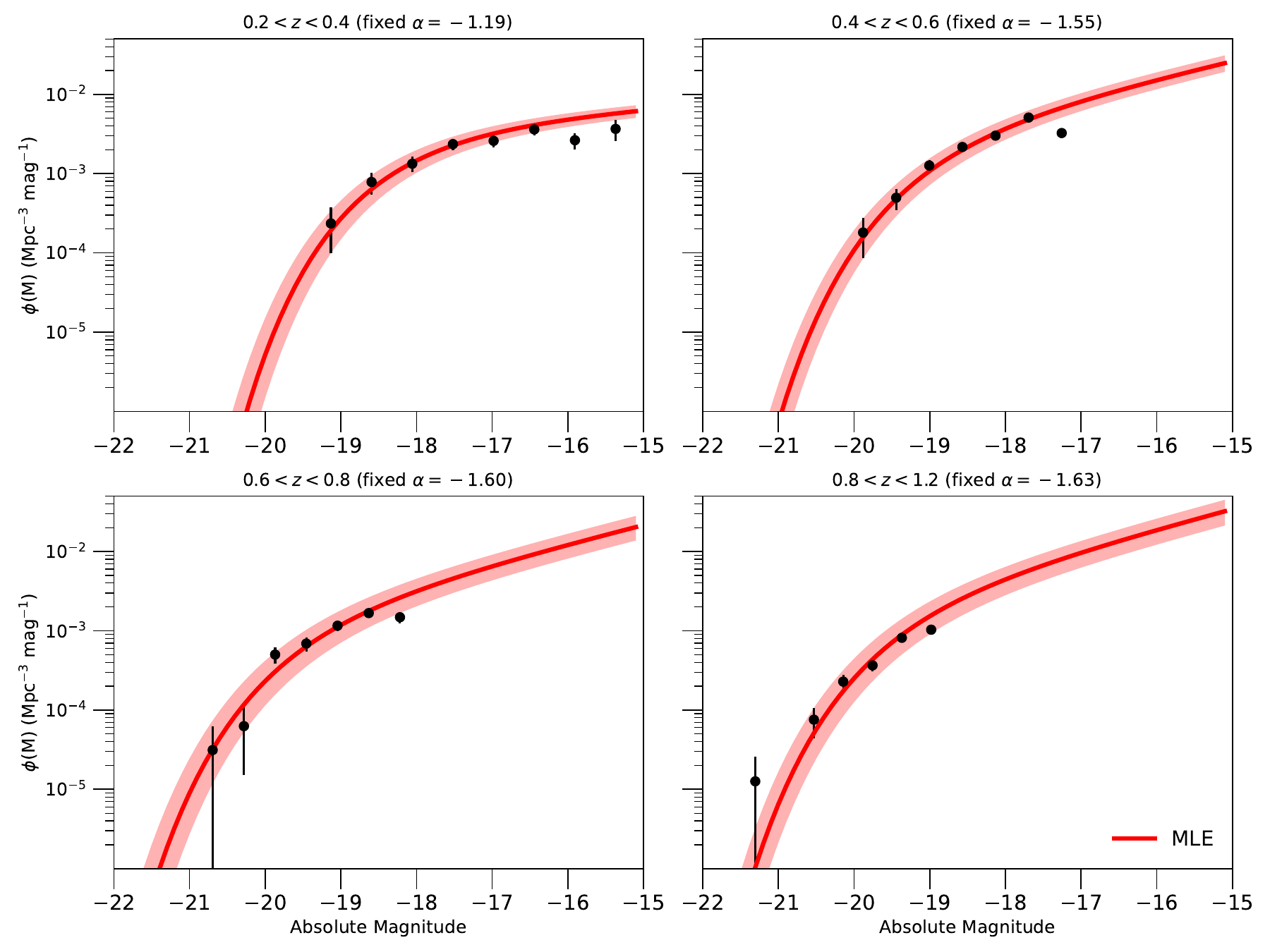}
    \caption{The UV LF in four different redshift bins as determined using UVOT observations of GOODS-N. The black dots are from the $V_\text{max}$ method {with errors from bootstrap resampling}. {The faintest bins are only comprised of a handful of galaxies due to incompleteness. As a result, the bootstrap error estimates are likely an underestimate of the true error.}  The solid red line is the MLE best fit Schechter function with the shaded region showing the $1\sigma$ uncertainty from $M^*$ and $\phi^*$. In each redshift bin, $\alpha$ is fixed using the values from \citet{2005ApJ...619L..43A}. We see clear evolution in $M^*$. The best fit LF is integrated to calculate the UV luminosity density which is converted to a star formation rate density using the scaling relation from \citet{1998ARA&A..36..189K}. }
    \label{fig:LF_bin}   
\end{figure*}

The UV LF as determined using MLE is shown in Figure~\ref{fig:LF_bin} for each redshift bins, {with the discrete $V_\text{max}$ points as a reference. We note that there is a slight turn down in the faintest bins. Specifically in the lowest redshift bin, this is likely related to the extended source discussion in \S~\ref{subsec:bias}, where we may be overestimating the completeness in the faintest magnitude bins. Additionally, the lowest absolute magnitude bins only contain a few objects. As a result, the bootstrap error is an underestimate}. The shaded region is determined using the one $\sigma$ errors for $M^*$ and $\phi^*$. 

\begin{deluxetable*}{c|cccc}
\tablecaption{Best fit Schechter function parameters \label{tab:params}}
\tablewidth{\textwidth}
\tablecolumns{6}
\tablehead{\colhead{Redshift range} & \colhead{$M^*$} & \colhead{$\phi^*/10^{-3}$ } & \colhead{$\alpha$ \tablenotemark{a}} & \colhead{SFRD$_\text{uncorr} / 10^{-2}$} \\
\colhead{} & \colhead{(Mag)} & \colhead{(Mpc$^{-3}$)} & \colhead{} & \colhead{($M_\odot$ yr$^{-1}$ Mpc$^{-3}$)}}
\startdata
Full sample ($V_\text{max}$) & $-19.05 \pm 0.11$ & $2.90 \pm 0.41$ & $-1.086 \pm 0.074$ & --- \\
Full sample (MLE) & $-19.04 \pm 0.11$ & $3.21 \pm 0.48$ & $-1.267 \pm 0.071$ & --- \\\hline
0.2-0.4 (free $\alpha$) & $-18.25 \pm 0.41$ & $3.34 \pm 1.47$ & $-1.31 \pm 0.20$ &  --- \\
0.4-0.6 (free $\alpha$) & $-18.66 \pm 0.29$ & $5.59 \pm 2.21$ & $-1.40 \pm 0.23$ &  ---  \\\hline
0.2-0.4 (fixed $\alpha$) & $-18.00 \pm 0.17$ & $4.29 \pm 0.61$ & $-1.19$ & $0.65 \substack{+0.22 \\ -0.17}$ \\
0.4-0.6 (fixed $\alpha$) & $-18.81 \pm 0.14$ & $4.23 \pm 0.69$ & $-1.55$ & $1.94 \substack{+0.62 \\ -0.51}$ \\
0.6-0.8 (fixed $\alpha$) & $-19.37 \pm 0.18$ & $2.13 \pm 0.59$ & $-1.6$ & $1.77 \substack{+0.83 \\ -0.64}$ \\
0.8-1.2 (fixed $\alpha$) & $-19.23 \pm 0.14$ & $3.26 \pm 0.92$ & $-1.63$ & $2.53 \substack{+1.16 \\ -0.94}$  \\
\enddata 
\tablenotetext{a}{Fixed values for $\alpha$ are taken from \citet{2005ApJ...619L..43A}.}
\end{deluxetable*}

\subsection{SFRD}

With Schechter function fits in various redshift bins, the star formation rate density as a function of redshift is calculated. First, the Schechter function must be integrated to get the luminosity density. In terms of the Schechter parameters, the luminosity density is given by:

\begin{align*}
    \rho_\text{UV} = \int_0^\infty L\ \phi(L) dL &= \phi^*\ L^*\ \Gamma(2+\alpha), 
\end{align*}
where $\Gamma$ is the gamma function. As $\alpha$ becomes steeper, small changes in $\alpha$ can greatly affect the luminosity density. For the extremes in $\alpha$ here, the ratio of $\Gamma(2-1.6)/\Gamma(2-1.19)$ is roughly 2. A non-zero lower limit on the integral changes the gamma function to the incomplete gamma function. {A lower limit of 0.03 $L^*$ is frequently assumed and is adopted here.}

This luminosity density can then be corrected for the effects of dust and converted into a star formation rate density. However, several simplifying assumptions must be made. First, the shape of the dust attenuation law is known to {vary between different galaxies \citep{2018ApJ...859...11S}, within a single galaxy \citep{2019MNRAS.486..743D}, and as a function of redshift \citep{2020MNRAS.496.5341B, 2024arXiv240205996M}.} A common approach to estimate the amount of attenuation is to use IRX-$\beta$ relationship from \citet{1999ApJ...521...64M}, where $\beta$ is the UV spectral slope $F_\lambda \propto \lambda^\beta$. However, recent works have found evidence that there is significant scatter in the IRX-$\beta$ relationship due to variations in the underlying dust attenuation curves \citep{2019ApJ...872...23S}. Work in dust attenuation have probed whether the observed attenuation is correlated with galaxy properties like optical opacity, metallicity, and inclination angle \citep{2018ApJ...859...11S, 2020ApJ...899..117S, 2017ApJ...851...90B}. 

Typically a $A_{FUV}-\beta$ is assumed to correct the luminosity density despite the uncertainty. \citet{2019ApJ...872...23S} proposed introducing additional terms to the $A_{FUV}-\beta$ relationship to reduce the impact of scatter, specifically recommending using stellar mass or Balmer optical depth. However, stellar mass estimates require the use of another, possibly uncertain, scaling relation or spectral energy distribution fitting, while Balmer optical depth requires spectra. {Additionally, there is uncertainty in how well the nebular attenuation, parameterized by the Balmer decrement, traces the UV continuum attenuation as line emission and stellar emission are distinct. The line emission is due to the most massive stars embedded in birth clouds, while the stellar UV emission comes from stars that are only attenuated by the ambient ISM \citep{2000ApJ...539..718C, 2018ApJ...859...11S}.}

The scaling relation to go from luminosity is more well understood but the exact calibration can vary. Within the scaling relation used are assumptions about the initial mass function (IMF) and star formation history (SFH). Whether or not the IMF is universal is the cause of some debate \citep{2010ARA&A..48..339B}. The SFH is expected to vary from galaxy to galaxy based on several stochastic factors such as merger rates, available gas mass, and star formation efficiency. 

With these caveats, we use the relation from \citet{1999ApJ...521...64M},
\begin{align*}
    A_{FUV} = 4.43 + 1.99 \beta,
\end{align*} 
where $\beta$ is the observed UV slope calculated between 1250 and 2600 \AA, to correct for dust. {This slope is historically measured using spectra. However, $\beta$ measured from photometric colors became common due to the ubiquity of GALEX data \citep{2004MNRAS.349..769K}. We estimate $\beta$ from a power law fit to photometry corresponding to a rest frame wavelength between 1250 and 2800 \AA.} In addition to the canonical relation from \citet{1999ApJ...521...64M}, we also explore the $A_{FUV}-\beta$ relationship from \citet{2011ApJ...726L...7O} which uses the same sample and improved aperture matching and finds less attenuation for a fixed $\beta$. Other empirically measured IRX-$\beta$ relationships are given in \citet{2012ApJ...755..144T} and \citet{2014ApJ...796...95C}. 

We use the SFR scaling relation from \citet{1998ARA&A..36..189K}, 
\begin{align*}
    \text{SFR(UV)} = 1.4 \times 10^{-28} L_{\text{UV}}.
\end{align*} 
The updated SFR indicator from \citet{2012ARA&A..50..531K}, based on \citet{2011ApJ...741..124H} and \citet{2011ApJ...737...67M}, will be discussed later. 

As the attenuation correction is highly uncertain, the uncorrected (observed) SFRD is typically reported and is given in Table~\ref{tab:params} for each redshift bin. We will discuss how these chosen scaling relations affect the interpretation of the results in Section~\ref{sec:dust_atten}.

\subsection{Redshift Evolution}

With our sample binned by redshift, we are able to explore redshift evolution of the galaxy population over a large fraction of the age of the Universe. 

\begin{figure}
    \centering
    \includegraphics[width=0.5\textwidth]{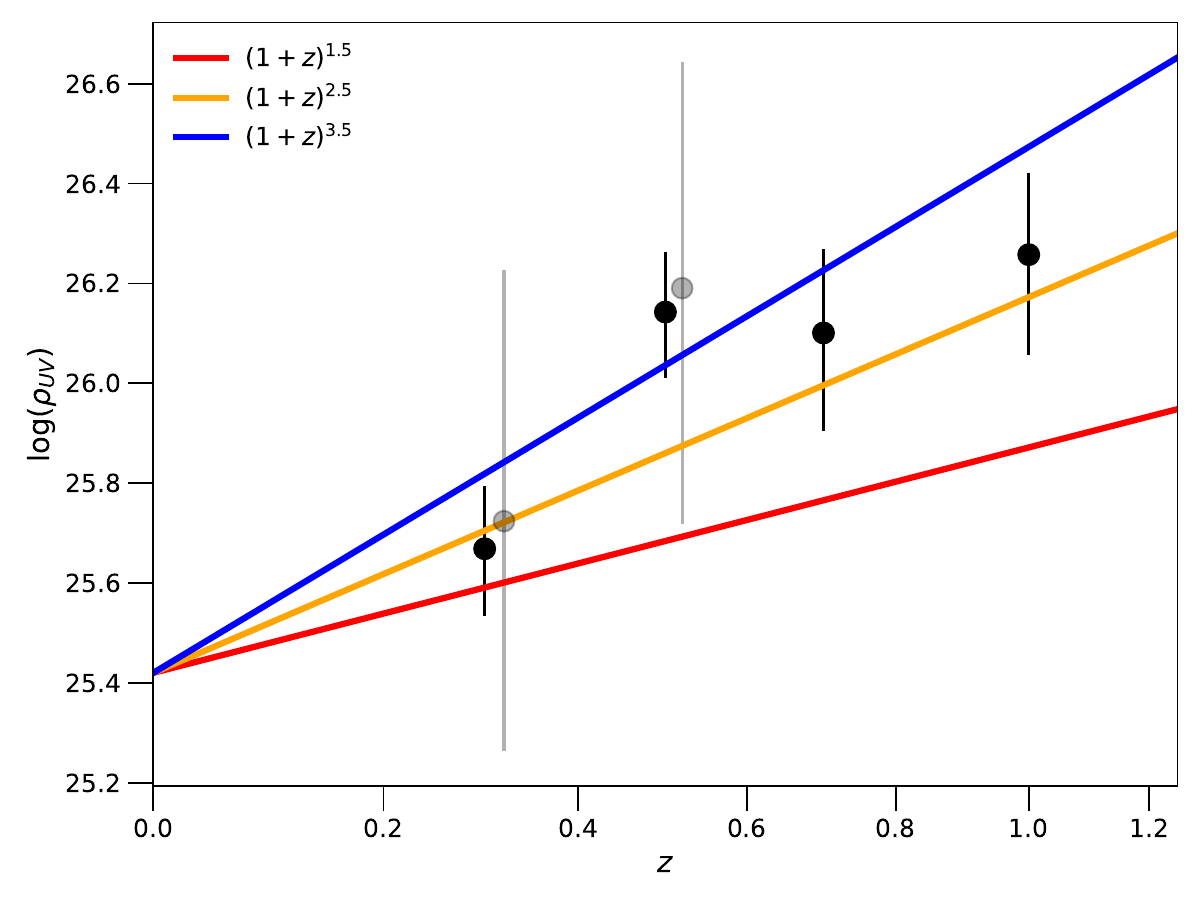}
    \caption{The observed UV luminosity density $\log(\rho_\text{UV})$ as a function of redshift. The grey points are from the MLE fits with $\alpha$ as a free parameter, while the black points are with $\alpha$ fixed using the values in \citet{2005ApJ...619L..43A}. {We fit a power law evolution model of the form $\log(\rho_\text{UV}) \propto (1+z)^n$. The best fit model is $n=3.04 \pm 1.38$. Using the intercept from the fit, we plot different evolution models corresponding to $n=$1.5 (red), 2.5 (yellow), 3.5 (blue).}}
    \label{fig:z_evo}
\end{figure}

Figure~\ref{fig:z_evo} shows {the luminosity density in the four different redshift bins as well as lines showing different rates of evolution}. We fit a power law and find that the observed luminosity density evolves as $(1+z)^{3.04 \pm 1.38}$ with the observed luminosity density $\log(\rho)=25.44 \pm 0.27$ at $z\sim0$. {This redshift evolution is consistent with the evolution reported in \citet{2005ApJ...619L..47S} and \citet{2015ApJ...808..178H}. \citet{2005ApJ...619L..15W} calculated the observed luminosity density and found a value of $\log(\rho_{UV}) = 25.55 \pm 0.12$, consistent with our results.} This Figure also shows how our values for the UV luminosity density are consistent between our fits whether $\alpha$ is fixed (solid black) or free (grey). However, due to the degeneracies between the LF parameters, the errors are much larger when $\alpha$ is left as a free parameter.  

\section{Discussion}
\label{sec:disc}

\subsection{Comparison to Literature}

\subsubsection{Number Counts}
As in \citet{2009ApJ...705.1462H}, we have shown the utility of using repeat UVOT observations of the same field to get deep photometry of star-forming galaxies at $z<1.5$. Our number counts agree with past literature results from GALEX NUV \citep{2005ApJ...619L..11X} and \swift/UVOT \citep{2009ApJ...705.1462H} and are qualitatively similar to the FUV number counts from HST \citep{2006AJ....132..853T}.

UVOT occupies a unique niche between GALEX and HST. While UVOT does not go as deep as HST or as wide as GALEX, it can achieve depths of $m_{AB}\sim24$ in short observations of a few kiloseconds per day taken over the course of a year in a larger area than could be observed by HST in a single pointing. This approach has the advantage of ease of scheduling as the short daily observations are spread over a long time frame, less susceptibility to cosmic variance, and moderate depth possible to get scarce UV data of galaxies at $z\lesssim1.5$.

\subsubsection{Luminosity Function}

We calculate the best fit LF of our entire sample as well as in four distinct redshift bins. Our best fit LF in each redshift bin shows evolution in $M^*$ consistent with other works. \citet{2015ApJ...808..178H}, which also used UVOT data to determine the UV LF in CDF-S, could not constrain the faint end slope $\alpha$. Here, we are able to constrain $\alpha$ but only for the two lowest redshift bins. For ease of comparison, we fix $\alpha$ using the values from \citet{2005ApJ...619L..43A} like in \citet{2015ApJ...808..178H}. Measurements of $\alpha$ show large scatter around $\alpha \sim -1.5$ between $0.5<z<2$ \citep{2010ApJ...725L.150O, 2016ApJ...832...56A} making it difficult to quantify its evolution. 

From our best fit UV LF, we find less luminous characteristic magnitudes compared to \citet{2015ApJ...808..178H}. The very luminous values in \citet{2015ApJ...808..178H} was noted by \citet{2022MNRAS.511.4882S}. They posited that the discrepancy could be the result of how galaxies were selected. \citet{2015ApJ...808..178H} used the SExtractor parameters from \citet{2009ApJ...705.1462H} to create their sample, specifically detecting sources with the $u$ band image. Here, we use SExtractor in a two phase approach and utilize a meta-image for detections made from the summed UVW2, UVM2, and UVW1 images. We specifically exclude the $u$ band to minimize the effects of the higher background. However, by excluding the $u$ band, we may be missing red sources or those in the highest redshift bin not detected by the blue filters.

The quoted 50\% completeness level in \citet{2015ApJ...808..178H} is $m_{AB}=24.1$ for $u$ band in roughly 125 ks. Here, using a different detection methodology we find 50\% completeness down to 23.85 mag in only 73 ks. The differences in $M^*$ is unlikely to be due to cosmic variance as \citet{2022MNRAS.511.4882S} using XMM-Newton OM data of the CDF-S field found less luminous characteristic magnitudes that agree with our results here. 

The differences in our detection method compared to \citet{2015ApJ...808..178H} could also explain our ability to determine the faint end slope in the two lowest redshift bins. While the previous work by \citet{2015ApJ...808..178H} had deeper data than what is presented here, with our different source selection methodology, we are able to achieve similar depths and constrain $\alpha$ in the two lowest redshift bins.

Additionally, \citet{2021MNRAS.506..473P} raised the issue of AGN contamination skewing the UV LF. Here, we do not see a measurable difference in $M^*$ when including X-ray detected sources. However, we agree that the non-thermal emission from AGN can significantly affect the UV LF and bias conclusions made about the star-forming galaxy population and therefore, we have excluded them.

When looking at the UV luminosity density, we find the expected evolution over the redshift range as shown in Figure~\ref{fig:z_evo}. \citet{2005ApJ...619L..47S} using GALEX data finds the UV luminosity density evolves as $(1+z)^{2.5\pm0.7}$ at $z<1$; for $z>1$, they find the power law index to be $(0.5\pm0.4)$. This is compared to our evolution which goes as $(1+z)^{\sim3}$. \citet{2005ApJ...619L..47S} has a similar number of galaxies to our work but over a larger area due to the source confusion limitation of GALEX. Our evolution is also faster than the values found by \citet{2015ApJ...808..178H} but do agree within the $1\sigma$ errors ($1.88 \pm 1.32$). 

\citet{2005ApJ...619L..47S} explored the UV luminosity density evolution of a subset of UV luminous galaxies ($M<-19.32$), finding more rapid evolution, $(1+z)^5$. {Our sample contains only a few} of these very UV luminous galaxies so a direct comparison is difficult. For the two higher redshift bins, the luminosity density values found here agree with the values reported in \citet{2022arXiv221200215S}. 

\begin{figure}
    \centering
    \includegraphics[width=0.47\textwidth]{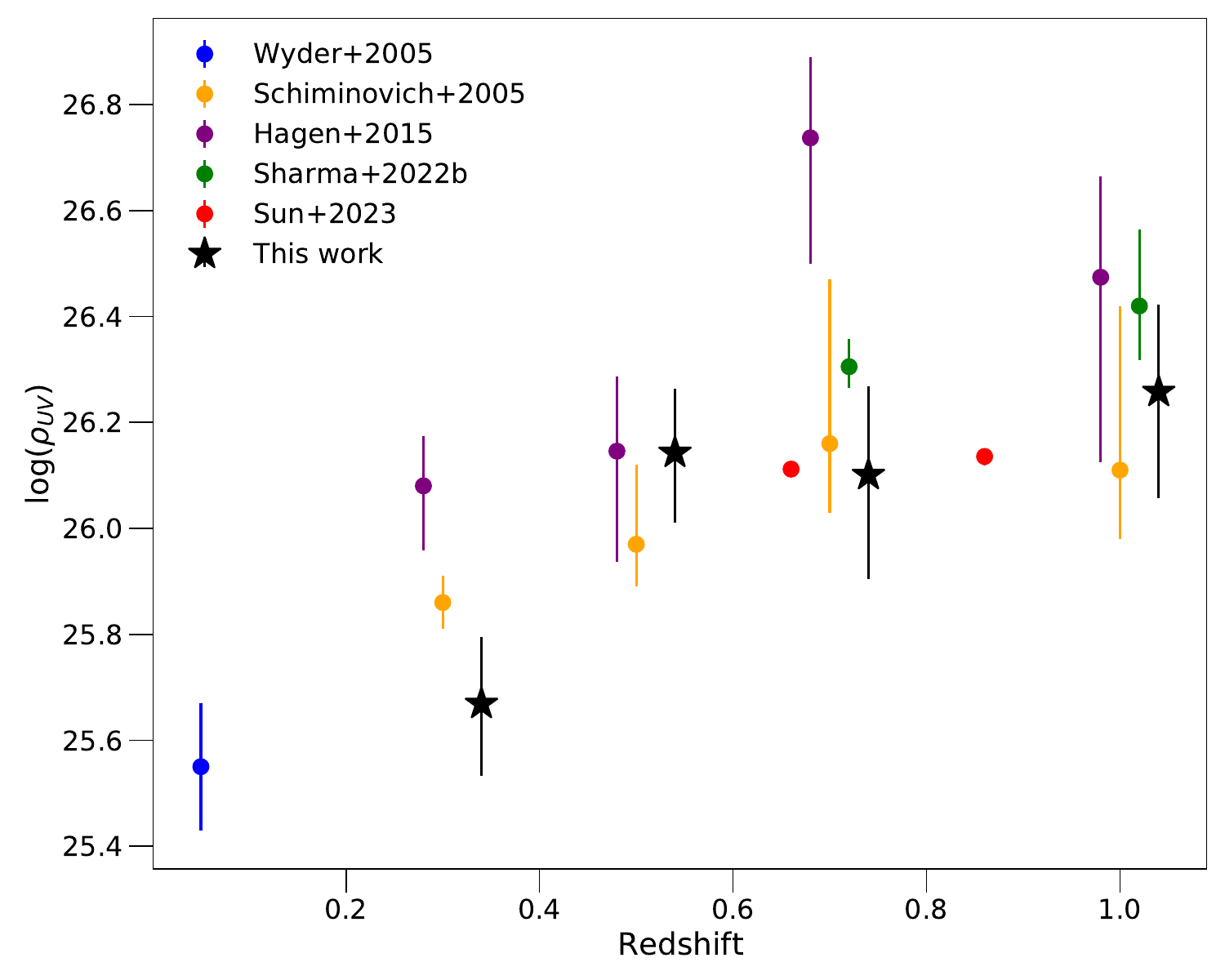}
    \caption{Our observed luminosity density (black stars) as a function of redshift compared to \citet{2005ApJ...619L..15W} (blue), \citet{2005ApJ...619L..47S} (yellow), \citet{2015ApJ...808..178H} (purple), \citet{2022arXiv221200215S} (green), and \citet{2023arXiv231115664S} (red). We see good agreement across all redshift bins. We focus on observed UV luminosity density as the chosen dust attenuation correction and SFR calibrations differ.}
    \label{fig:lit_comp}
\end{figure}

We compare the observed luminosity density found here with recent results in the literature and early results from GALEX. Figure~\ref{fig:lit_comp} shows the observed luminosity density from a variety of studies compared to the results here. We see relatively good agreement across the entire redshift range. However, scatter is present as each studies observes a different area and location on the sky. A larger analysis combining these results could provide tighter constraints on the global UV luminosity density evolution over this redshift range.

{We also test how the redshift bins affect the derived evolution. To do so, we use finer redshift bins (width of 0.1 from $z=0.2$ to $z=0.7$, 0.2 from $z=0.7-0.9$, and 0.3 between $z=0.9-1.2$. We again fix $\alpha$ using the values from \citet{2005ApJ...619L..43A}. We find that the uncorrected UV luminosity density evolves with a power law index of $3.21 \pm 0.96$ with this binning scheme, agreeing with the evolution determined above.}

\subsection{UV SF Scaling Relations}

Understanding the redshift evolution of the UV LF is of great interest as it is a direct tracer of star formation over cosmic time. However, in addition to systematics like cosmic variance, it is not necessarily straightforward to correct for the effects of dust attenuation and to convert from a UV luminosity to a SFR.    

The scaling relation we used for calculating SFR in Section~\ref{sec:results} is the canonical result from \citet{1998ARA&A..36..189K}. However, more recent calibrations exist from \citet{2011ApJ...741..124H} and \citet{2011ApJ...737...67M}. Additionally, \citet{2015ApJ...808..109M} provides a SFR calibration using local volume dwarf galaxies. The K98 scaling relation is intermediate between the more recent results. Specifically, \citet{2011ApJ...741..124H} gives SFRs that are 68\% of those from K98 and the \citet{2015ApJ...808..109M} values are $1.46\times$ larger than K98. This factor of two discrepancy between the extremes is a severe limitation in determining the normalization for the changing SFRD with redshift.

To better constrain SFR, composite SFR indicators are commonly used as monochormatic SFR indicators are much less reliable. By combining the UV and total IR emission or possibly emission line diagnostics, the SFR can be determined more precisely \citep{2013seg..book..419C, 2012ARA&A..50..531K, 2014ARA&A..52..415M}. However, practically this can be an issue as observations at UV and IR wavelengths require observations from different facilities. These multiwavelength observations may not exist for the field of interest.  

\citet{2019ApJ...877..140L} showed that SED fitting of 5000+ galaxies at $0.5<z<2.5$ produced SFRs that were 0.1 to 1 dex lower than the standard UV+IR indicators. They explain the discrepancy as due to the emission of stars older than 100 Myr. Performing SED fitting on this sample of galaxies would be an interesting test of the accuracy of the standard SFR scaling relations. 

\subsection{Impact of Assumed Dust Attenuation Correction}
\label{sec:dust_atten}

In addition to the uncertainty in the normalization of the SFRD caused by the SFR indicator chosen, correcting for the effects of internal dust attenuation can affect the normalization and evolution of the SFRD. The effects of dust attenuation are highly uncertain; a variety of dust attenuation curves have been observed in galaxies, with the biggest discrepancies and highest uncertainties in the UV. The empirical IRX-$\beta$ relationship as defined in \citet{1999ApJ...521...64M} is typically used to estimate the attenuation in the FUV band from the observed UV spectral slope. 

At it simplest, the attenuation law has been seen to vary between a shallow curve such as the starburst law reported in \citet{2000ApJ...533..682C} and the SMC extinction law. At 1500 \AA, these curves vary drastically. In terms of $A_V$, the attenuation at 5500 \AA, the FUV attenuation from the SMC curve is $4.77 A_V$ compared to $2.55 A_V$ for the Calzetti law. For a dusty galaxy with $A_V=1$, these corrections differ by more than 2 mag, almost an order of magnitude difference in luminosity. 

To correct our sample here for dust, we use the median UV slope $\beta$ to calculate $A_\text{FUV}$ for each redshift bin. $\beta$ was originally defined as a fit to 10 wavelength windows from UV spectra \citep{calzetti_94}. Robustly calculating it from photometry alone is challenging. Often, at low redshift, $\beta$ is estimated using the GALEX FUV-NUV color \citep{2004MNRAS.349..769K}. While the \citet{calzetti_94} windows range from 1250-2600 \AA, some works use photometry out to 3000 \AA\ which can affect the fit. The different methodologies used to calculate $\beta$ can complicate comparing across studies.

{For each galaxy, a power law was fit to photometry corresponding to a rest frame wavelength between 1250 and 2600 \AA\ in order to estimate $\beta$. For every redshift bin, we find the median $\beta$ to be roughly $-1.2$ but with significant scatter, where $\beta$ ranges from -2.5 to +1. \citet{2020MNRAS.496.5341B} predicts that the FUV attenuation, which is often estimated using $\beta$, should increase rapidly up to $z\sim2$. \citet{2014ApJ...793L...5K} using HST data calculated $\beta$ for galaxies across the redshift range $1<z<8$, finding that the highest redshift galaxies were bluer and that the redshift evolution was strongest in the faintest galaxies.}

\begin{figure*}
    \centering
    \includegraphics[width=0.9\textwidth]{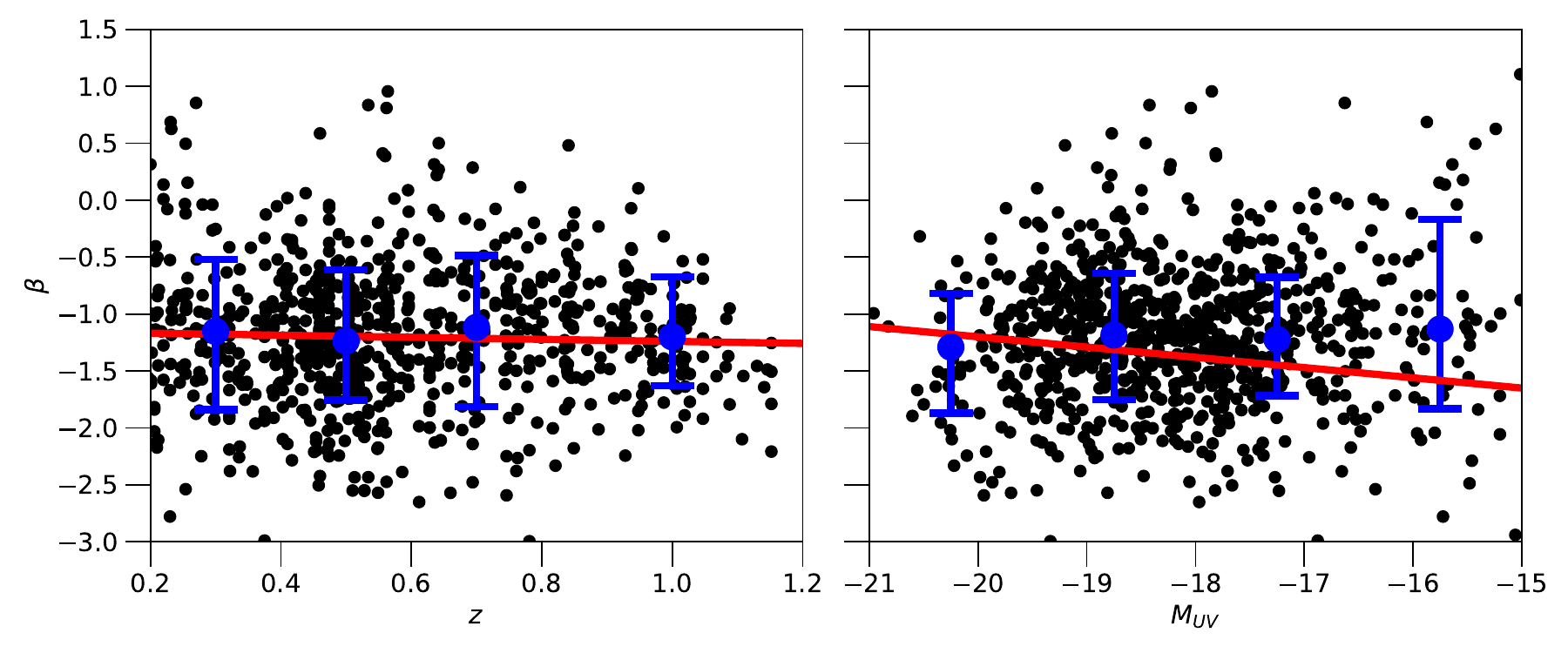}
    \caption{{The UV spectral slope $\beta$ as a function of absolute UV magnitude and redshift. We plot the measured UV slopes of each galaxy used in calculating the LF (black points), the median value in a given bin {(blue points)}, and the 16th and 84th percentiles {(blue errorbars)}. The red line shows the change in $\beta$ with respect to $M_\text{UV}$ and $z$ found in \citet{2014ApJ...793L...5K}, {extrapolated the $z\sim1$ down to $z\sim0$}. Here, {we find} no strong trend at $z<1$ but the observed relationship {from \citet{2014ApJ...793L...5K} is} not ruled out due to the large scatter in $\beta$. {We investigate further by investigating the redshift trend (left) in smaller magnitude bins and find no trend. Similarly, we bin the magnitude relation (right) in smaller time bins and again find no trend.} \label{fig:beta_v_L}}}
\end{figure*}

{Figure~\ref{fig:beta_v_L} shows $\beta$ versus absolute UV magnitude and redshift for this sample. Also shown are the median value and $1\sigma$ error in each bin and the evolution in $\beta$ with respect to $M_\text{UV}$ or $z$ reported in \citet{2014ApJ...793L...5K}, {extrapolated from $z\sim1$}. In terms of $\beta$ and by extension, $A_\text{FUV}$, we see no obvious evolution with redshift, contrary to some predictions in the literature. {We also see no noticable trend with absolute magnitude}. Due to the large scatter {in $\beta$}, our values are not in tension with $\beta$ evolving as a function of $M_\text{UV}$ and $z$. {It is also possible that} due to our selection methodology, we are missing many highly attenuated sources.} {Regardless, it is clear that the large scatter in $\beta$ is more significant than any evolution with redshift or magnitude.}

{The median $A_\text{FUV}$ in each redshift bin was determined using the \citet{1999ApJ...521...64M} relationship. However, the reanalysis of the original starburst sample by \citet{2011ApJ...726L...7O} finds the attenuation is roughly 0.5 mag less after the improved aperture matching. Because of these differences, the IRX-$\beta$ calibration assumed is another uncertainty that can greatly affect the normalization of the dust-corrected SFRD. }

Using the FUV attenuation corresponding to the median $\beta$ in each redshift bin, we find the corrected luminosity density scales as $(1+z)^{2.97 \pm 1.10}$ with $\log(rho)=26.28 \pm 0.22$ at $z=0$. This evolution is in line with the UV+IR evolution from \citet{2014ARA&A..52..415M}. {If we found that attenuation at $z\sim1$ was greater, the evolution in the dust-corrected SFRD would be much faster than the consensus value.}


{We investigate $\beta$ as a function of redshift or absolute magnitude by looking at studies of higher redshifts. \citet{2009ApJ...705..936B} find that more luminous galaxies are more attenuated and therefore have redder UV slopes for a sample of UV selected galaxies at $z\sim2.5$. Meanwhile, \citet{2008ApJS..175...48R, 2018ApJ...853...56R} found no correlation between the dust attenuation and observed UV luminosity at $z\sim2$. \citet{2012ApJ...756..164F} and \citet{2014ApJ...793..115B} find evidence of a correlation at $z\sim6$ and $z\sim4$, respectively.}

{Looking at other works at these low redshifts, \citet{2023ApJ...946...90M} found no correlation between UV magnitude and $\beta$ at $0.4<z<0.75$ using Astrosat UVIT. A similar study by \citet{2024MNRAS.528.1997S} investigated the dust properties of XMM-Newton OM UVW1 selected galaxies in CDF-S using a stacking analysis of Herschel PACS and SPIRE data. They find that IRX, a proxy for attenuation, is highest at low UV luminosities for galaxies at $z=0.6-1.2$. Prior GALEX work by \citet{2007ApJS..173..432X} found that for $z\sim0.6$ UV selected galaxies, IRX does not vary with $L_{UV}$ but that $z\sim0$ galaxies show a dependence where IRX is less at low UV luminosities. For our sample of UV selected galaxies at $z<1.2$, we find no correlation between $\beta$ and absolute UV magnitude.}

{There is a lack of consensus, especially at low redshift, as to whether UV luminosity and attenuation are correlated. Selection effects are also a concern. As this sample is selected solely using the three near-UV filters of UVOT, highly obscured sources may be missed. These sources would have both low observed luminosities and red UV slopes. Additionally, the low redshift bins probe a relatively small volume, so cosmic variance could explain the paucity of red, luminous sources.}

{As mentioned, some studies such as \citet{2012ApJ...756..164F} and \citet{2014ApJ...793..115B} see a $\beta-L_\text{UV}$ relationship at high redshift, where low luminosity galaxies are bluer. As strong evidence for a relationship is not seen in studies of the low redshifts, the lack of a trend at low $z$ could be due to changes over cosmic time, specifically evolution in the interactions between the massive UV-emitting stars and the ISM. For young galaxies, dust is highly concentrated in star-forming regions, where the UV attenuation will be strongest. Later, through multiple generations of star formation, stellar feedback mixes the ISM, causing the dust to become diffuse across the entire galaxy rather than located near the site of star formation. Along with this different star-dust geometry at low redshift, age and metallicity differences become more apparent, and in turn will affect the color of galaxies. Any trend between UV emission from massive star formation and dust attenuation may be obscured as the star-dust geometry becomes more complex and age and metallicity become important factors.}

This is seen in \citet{2018ApJ...869...70N}, which investigated the physical cause of dust attenuation law variations using cosmological simulations. They assume a single underlying extinction law and model the effects that geometry and radiative transfer have on the net attenuation law over cosmological time. They find that at $z\sim6$, the attenuation law shows only small variations due to galaxies at this redshift having relatively uniform stellar ages and less dispersion in the star-dust geometry. However as galaxies evolve, the dispersion in attenuation curve shapes at $z\sim0$ becomes substantial. 

{For this sample of UV selected galaxies at $0.2\leq z\leq 1.2$, $\beta$ is relatively constant with respect to both redshift and absolute magnitude. This shows that, specifically at lower redshifts, dust attenuation is not a simple function of redshift or luminosity. The scatter is driven primarily by differences between galaxies. The lack of obvious redshift evolution in $\beta$ and consequently FUV attenuation is likely due to the great variation in dust attenuation between galaxies at low redshift.}

{Another complication is the various different empirical IRX-$\beta$ relationships reported in the literature. For starburst galaxies, there is little scatter relative to the \citet{1999ApJ...521...64M} curve. However, for normal star-forming galaxies, there is significant scatter.} \citet{2018MNRAS.474.1718N} explores the origin of the scatter seen in the IRX-$\beta$ relationship. They show that the age of the stellar population, the geometry of the stars and dust, and the underlying extinction law can all shift the IRX-$\beta$ relation away from the standard curve from \citet{1999ApJ...521...64M}. {By detailed modeling of the stellar populations via SED fitting, the attenuation curve can be estimated. Otherwise, there will be uncertainty in the efficacy of using the standard \citet{1999ApJ...521...64M} correction, as it corresponds to a \citet{2000ApJ...533..682C} attenuation law.}

Ideally, as we have UV slope information for each galaxy in our sample, we could calculate the UV LF using the dust corrected luminosities. This would be preferable as the individual galaxies that comprise the sample can have varying attenuation curves. In terms of the Schechter parameters, {correcting for dust prior to fitting the UV LF} would have the largest effect on $L^*$ but could also have a small effect on the faint end slope $\alpha$ as galaxies observed as UV faint could simply be highly obscured. For a fixed $L^*$ and $\phi^*$, a flattening of the faint end slope would lead to a smaller luminosity density and SFRD. However, in practice it is challenging to accurately account for things like completeness in calculating the UV LF after correcting individual galaxies for dust. 


\section{Conclusions}
\label{sec:conclusion}

In this work, we present deep ultraviolet observations of the GOODS-North field using the UVOT aboard the Neil Gehrels Swift Observatory. These data complement previous UVOT observations of CDF-S as well as the extensive multiwavelength observations of GOODS-N in the literature. We catalog all detected sources, provide color information from the four filters spanning the near UV, and cross-match our objects with \citet{2014ApJS..215...27Y} to construct SEDs ranging from the UV to IR.

We report a catalog of 1000+ sources identified in at least one of four UV bands. The UV galaxy number counts as a function of apparent magnitude are calculated and in good agreement with previous results from GALEX, Swift/UVOT and HST. After cross-matching our catalog with \citet{2014ApJS..215...27Y}, we perform K-corrections to our UV data to get the absolute FUV magnitudes of the detected galaxies. Our main results are as follows:

\begin{enumerate} 
    \item Using both $V_\text{max}$ and MLE methods, we fit the LF with a Schechter function in four different redshift bins between $0.2<z<1.2$. We find evidence for evolution of $M^*$, consistent with other works. Compared to prior UVOT observations of CDF-S, we find fainter characteristic luminosities and are able to constrain the faint-end slope $\alpha$ in the two lowest redshift bins. For ease of comparison with the literature, we fix $\alpha$ using the GALEX results from \citet{2005ApJ...619L..43A}. 
    \item Using our best fit LFs, we calculate the luminosity density and star formation rate density evolution. Using the color information provided by UVOT, we calculate $\beta$ from the observed photometry and use the IRX-$\beta$ relation from \citet{1999ApJ...521...64M} to correct for the effect of dust attenuation. Our dust corrected SFRD agrees with previous results. 
    \item We discuss that different star formation rate scaling relations can change the SFRD normalization by a factor of two. Additionally, {we discuss the difficulty in correcting for dust attenuation, which can have a large impact on the evolution and normalization of the corrected SFRD. In particular, we highlight the large observed spread in $\beta$ seen for galaxies in this redshift range.} 
    \item {Compared to studies at high redshift, we do not find strong evidence that $\beta$ is correlated with UV luminosity or redshift. As a redshift and UV luminosity dependence is seen in UV selected samples of high redshift galaxies and not surveys at $z\sim1$ implies that galaxy to galaxy differences in dust attenuation at low redshift are stronger than any evolution in $\beta$ with respect to $L_\text{UV}$ or $z$, likely driven by the complex interplay between star formation and the ISM.} 
\end{enumerate}

While the field pushes towards studying large samples of high redshift objects, understanding the UV emission of low to intermediate redshift galaxies is important to fully track the evolution in star formation over cosmic time. {At these low redshifts, the differences between dust attenuation in these galaxies complicates studying the evolution in the star formation rate density.} Using the sample of galaxies provided here, further work is possible by modeling the full spectral energy distributions in order to better constrain dust attenuation and {compare SED fitting derived SFRs to the standard scaling relations}.

\section*{Acknowledgements}

{The authors wish to thank the anonymous referee for comments that greatly improved this manuscript.} We acknowledge sponsorship at the Pennsylvania State University by NASA contract NAS5-00136. This research was also supported by the NASA Astrophysics Data Analysis Program (ADAP) through grant NNX16AF35G. The Institute for Gravitation and the Cosmos is supported by the Eberly College of Science and the Office of the Senior Vice President for Research at the Pennsylvania State University. MJP acknowledges support from the UK Space Agency, grant number ST/X002055/1. We thank Gautam Nagaraj for conversations regarding processing UVOT deep field data. This research has made use of NASA's Astrophysics Data System. This research has made use of data and/or software provided by the High Energy Astrophysics Science Archive Research Center (HEASARC), which is a service of the Astrophysics Science Division at NASA/GSFC and the High Energy Astrophysics Division of the Smithsonian Astrophysical Observatory.

\software{Astropy \citep{astropy:2013, astropy:2018, astropy:2022}, \texttt{kcorrect} \citep[version 5.0.1b]{2007AJ....133..734B}, emcee \citep{2013PASP..125..306F}, corner \citep{2016JOSS....1...24F}. }

\bibliography{cdfn}{}
\bibliographystyle{aasjournal}

\end{document}